\documentclass[10pt,twocolumn,letterpaper]{article}

\usepackage{iccv}              

\definecolor{iccvblue}{rgb}{0.21,0.49,0.74}
\usepackage[pagebackref,breaklinks,colorlinks,allcolors=iccvblue]{hyperref}

\def\confYear{2025}

\usepackage{colortbl} 
\usepackage{algorithm}
\usepackage{algpseudocode}

\title{IntrinsicReal: Adapting IntrinsicAnything from Synthetic to Real Objects}

\author{
Xiaokang Wei\textsuperscript{1} \quad
Zizheng Yan\textsuperscript{2} \quad
Zhangyang Xiong\textsuperscript{4} \quad
Yiming Hao\textsuperscript{2}\\
Yipeng Qin\textsuperscript{3}\quad
Xiaoguang Han\textsuperscript{2}\thanks{Corresponding Author} \\
\\
\textsuperscript{1}The Hong Kong Polytechnic University \quad 
\textsuperscript{2}The Chinese University of Hong Kong, Shenzhen \quad \\
\textsuperscript{3}Cardiff University \quad 
\textsuperscript{4}NanJing XiaoZhuang University
}

\begin{document}

\twocolumn[{
\maketitle
\begin{center}
\vspace{-2mm}
\captionsetup{type=figure}
\includegraphics[width=\linewidth]{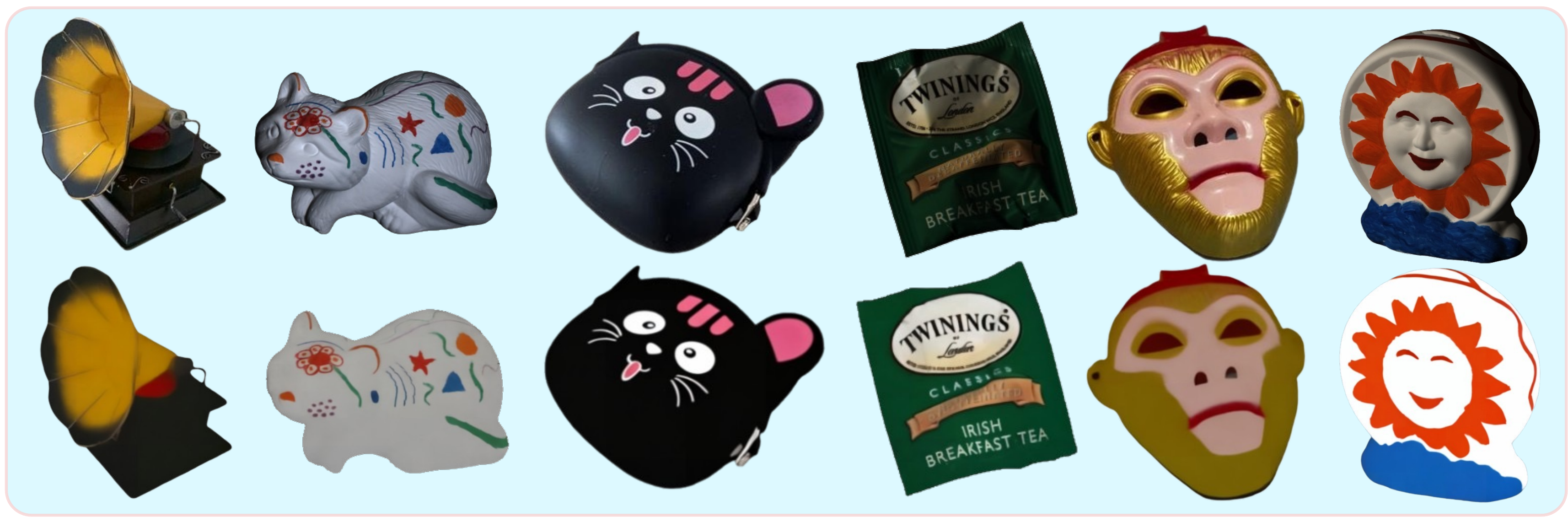}%
\vspace{-2mm}
\caption{Our IntrinsicReal demonstrates promising performance for intrinsic image decomposition of real-world object images. Top row: input images. Bottom row: predicted albedos.}
\label{fig:teaser}
\end{center}
}]

\begin{abstract}
Estimating albedo ({\it a.k.a., intrinsic image decomposition}) from single RGB images captured in real-world environments (\eg, the MVImgNet dataset) presents a significant challenge due to the absence of paired images and their ground truth albedos.
Therefore, while recent methods (\eg, IntrinsicAnything) have achieved breakthroughs by harnessing powerful diffusion priors, they remain predominantly trained on large-scale synthetic datasets (\eg, Objaverse) and applied directly to real-world RGB images, which ignores the large domain gap between synthetic and real-world data and leads to suboptimal generalization performance.
In this work, we address this gap by proposing {\bf IntrinsicReal}, a novel domain adaptation framework that bridges the above-mentioned domain gap for real-world intrinsic image decomposition.
Specifically, our IntrinsicReal adapts IntrinsicAnything to the real domain by fine-tuning it using its high-quality output albedos selected by a novel dual pseudo-labeling strategy: i) pseudo-labeling with an {\bf absolute} confidence threshold on classifier predictions, and ii) pseudo-labeling using the {\bf relative} preference ranking of classifier predictions for individual input objects.
This strategy is inspired by human evaluation, where identifying the highest-quality outputs is straightforward, but absolute scores become less reliable for sub-optimal cases. In these situations, relative comparisons of outputs become more accurate.
To implement this, we propose a novel two-phase pipeline that sequentially applies these pseudo-labeling techniques to effectively adapt IntrinsicAnything to the real domain. Experimental results show that our IntrinsicReal significantly outperforms existing methods, achieving state-of-the-art results for albedo estimation on both synthetic and real-world datasets.

\end{abstract}    
\section{Introduction}
\label{sec:intro}

\noindent
Intrinsic image decomposition is a central problem in computer vision, aiming to separate an image into its intrinsic components, such as albedo (reflectance) and shading (illumination), from a single RGB image. 
This task is fundamental for applications in inverse rendering and scene understanding, enabling advanced use cases in virtual and augmented reality by allowing the reconstruction and re-rendering of a scene's 3D structure from 2D images~\cite{mildenhall2021nerf}.
While recent advances in data-driven approaches have significantly improved intrinsic image decomposition~\cite{bell2014intrinsic}, a persistent challenge remains in bridging the domain gap between synthetic and real data.

\begin{figure}
\vspace{-0.2cm}
    \centering
    \includegraphics[width=0.98\linewidth]{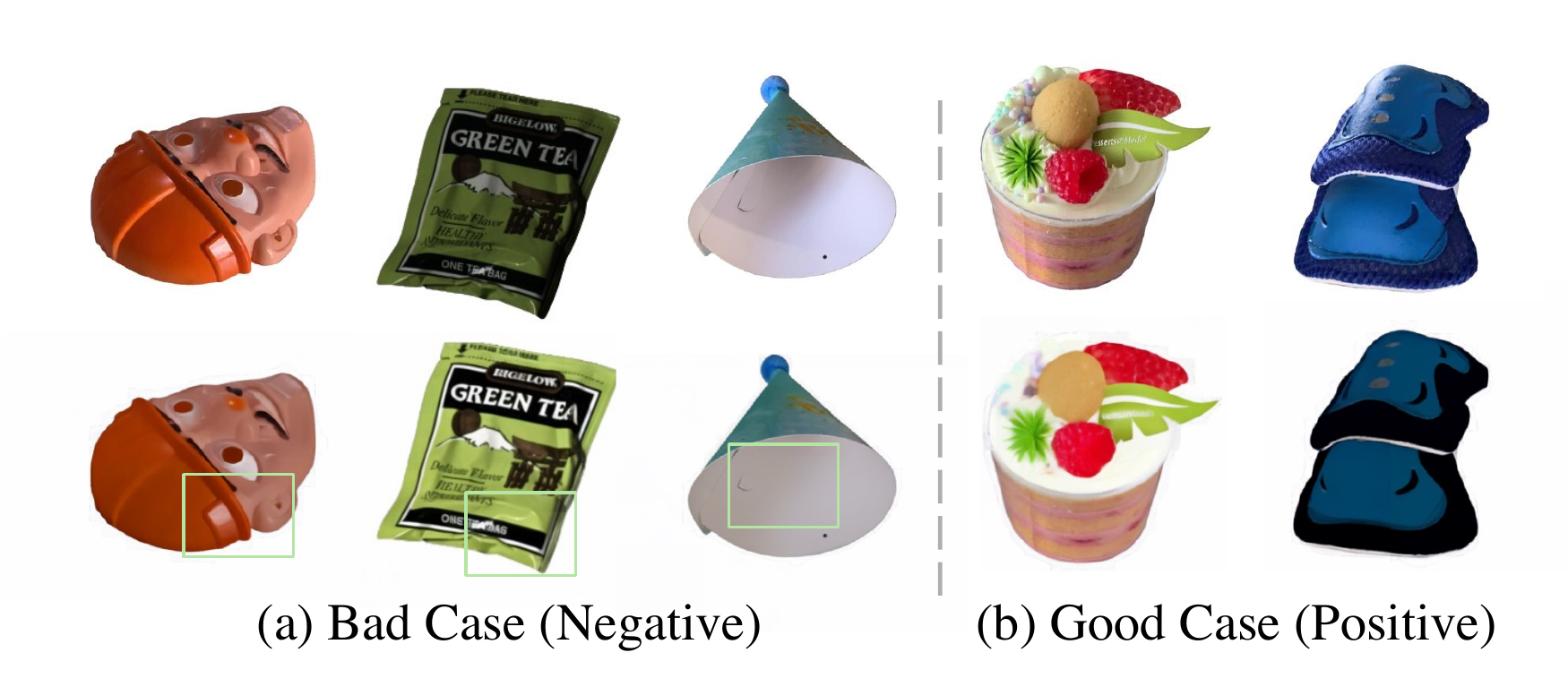}
    \vspace{-0.2cm}
    \caption{\textbf{Applying IntrinsicAnything~\cite{chen2024intrinsicanything} on MVImgNet.} 
    Top row: input images.
    Bottom row: predicted albedos. It can be observed that IntrinsicAnything fails under certain circumstances.}
    \label{fig:intrinsic_pred}
    \vspace{-0.4cm}
\end{figure}

Specifically, IntrinsicAnything~\cite{chen2024intrinsicanything} has shown impressive results in intrinsic image decomposition by leveraging a large-scale synthetic dataset, namely Objaverse~\cite{objaverse}. 
However, when applied directly to real-world RGB images, the model faces a substantial domain gap between synthetic and real data, resulting in suboptimal generalization performance, as demonstrated on Fig.~\ref{fig:intrinsic_pred}. 
Addressing this issue, a naive idea would be to train IntrinsicAnything on real-world data. 
However, this is infeasible as synthetic datasets provide paired albedo and shading information, whereas real-world datasets lack such ground-truth pairs. 
This domain gap severely limits the effectiveness of current models in accurately decomposing real-world images, emphasizing the need for novel strategies to bridge the synthetic-to-real domain gap for real-world intrinsic image decomposition.

In this work, we address the above-mentioned gap by proposing {\bf IntrinsicReal}, a novel framework that adapts synthetic-trained intrinsic decomposition models to real-world data.
In a nutshell, our IntrinsicReal finetunes IntrinsicAnything with its own high-quality output albedos that are selected by a novel dual pseudo-labeling strategy.
This strategy draws inspiration from human evaluation, where identifying the highest-quality outputs is straightforward, but absolute evaluation scores become less reliable for sub-optimal cases. In these situations, relative comparisons of outputs become more accurate.
Accordingly, given a classifier that assesses the quality of albedos, we create two types of pseudo-labels using i) an {\bf absolute} confidence threshold applied to the classifier predictions and ii) {\bf relative} preference rankings derived from different classifier predictions for individual input objects.
To leverage these pseudo-labels, we introduce a novel two-phase pipeline that sequentially applies these labeling techniques.
In Phase 1, we introduce a novel iterative joint updating scheme between the albedo generation model (referred to as {\it IntrinsicReal-Model}), and the classifier (denoted as {\it IntrinsicReal-Classifier}). This scheme consists of three stages:
i) {\it Initialization}: We initialize IntrinsicReal-Model as a synthetic-trained IntrinsicAnything model and use it to generate corresponding albedo images of those in the MVImgNet dataset. We then train an initial albedo-shading classifier (\ie, IntrinsicReal-Classifier) to distinguish between the generated albedo and RGB images.
Finally, we manually label a small set of albedo images generated by the IntrinsicReal-Model on real-world data, and categorize them into positive and negative sets, respectively.
ii) {\it Real-domain Adaptation}: This stage comprises four steps: 
First, we use the positive and negative sets to fine-tune the IntrinsicReal-Classifier. 
Next, we apply the IntrinsicReal-Model to real-world data to generate corresponding albedos, using the IntrinsicReal-Classifier to assign pseudo-labels. 
In the third step, we fine-tune IntrinsicAnything with these pseudo-labeled albedos as the updated IntrinsicReal-Model. Finally, we update the positive and negative sets based on the refined IntrinsicReal-Classifier and IntrinsicReal-Model outputs.
iii) {\it Iteratively Joint Updating}: 
We iteratively repeat these four steps, gradually adapting the IntrinsicReal-Classifier and IntrinsicReal-Model to real-world data.
In Phase 2, we first extract two distinct albedos of an object from the outputs of successive iterations of the IntrinsicReal-Model in Phase 1. 
Next, we determine their relative preference ranking using the scores predicted by IntrinsicReal-Classifier.
Finally, we optimize the IntrinsicReal-Model using Diffusion-DPO~\cite{wallace2024diffusion} based on the derived preference ranking.
Extensive experiments and analysis on both synthetic and real-world datasets demonstrate the effectiveness of our method.
Our contributions include:
\begin{itemize}
\item We introduce a novel synthetic-to-real intrinsic decomposition task that uses unpaired real-world data (RGB only) to adapt models trained on synthetic data (RGB, albedo) to real-world scenarios.
\item We propose IntrinsicReal, a novel framework that introduces a dual pseudo-labeling strategy including i) pseudo-labeling with an absolute confidence threshold applied to the classifier predictions; and ii) pseudo-labeling using relative preference rankings derived from the classifier predictions for individual input
objects; to select high-quality albedo outputs for fine-tuning IntrinsicAnything to the real domain.
\item We introduce a novel two-phase pipeline that sequentially applies the two pseudo-labeling techniques, including a novel iterative joint updating scheme and Direct Preference Optimation (DPO).
\item Extensive experiments and analysis on both synthetic and real-world datasets demonstrate the effectiveness of our IntrinsicReal framework.
\end{itemize}

\section{Related Work}\label{sec:related_work}

\subsection{Intrinsic Image Decomposition} 
Intrinsic image decomposition aims to separate surface reflectance from illumination effects within a single image, serving as a fundamental mid-level vision task essential to many inverse rendering pipelines. 
Classical optimization-based approaches \cite{bi20151, cheng2019non, li2014single} rely on hand-crafted priors, which perform well on small, controlled datasets but struggle with complex real-world objects \cite{kovacs2017shading}. Advances in physically-based rendering have enabled data-driven intrinsic image decomposition models trained on synthetic datasets, with deterministic methods such as CNN-based approaches \cite{baslamisli2018cnn, meka2018lime, shi2017learning} leveraging rendered datasets \cite{chang2015shapenet, butler2012naturalistic, grosse2009ground}. However, intrinsic image decomposition remains challenging due to its inherent ambiguity: the observed appearance of objects arises from complex interactions between lighting and materials, making it difficult to disentangle these factors as lighting effects could be easily baked in material attributes and vice versa.

Recently, the advent of diffusion models has opened new avenues for intrinsic image decomposition by exploiting their strong priors learned from large-scale data. 
For example, \cite{zeng2024rgb, kocsis2023intrinsic, chen2024intrinsicanything, litman2024materialfusion} leverage diffusion models to address the ill-posed nature of intrinsic image decomposition probabilistically, using conditional generation techniques trained on extensive datasets \cite{li2018cgintrinsics}. \cite{kocsis2023intrinsic} focuses on indoor images and highlights how CGI pipelines often embed lighting into albedo.
Rather than extending latent space channels as in \cite{kocsis2023intrinsic, litman2024materialfusion}, \cite{zeng2024rgb} introduces instruction prompts (\eg, ``normal'') to predict various intrinsic components with a fixed latent size, as well as building a unified bidirectional framework that links RGB and intrinsic channels. 
\cite{chen2024intrinsicanything} first integrates diffusion-based intrinsic image decomposition into an inverse rendering pipeline, using it to regularize optimization. However, these approaches primarily adapt diffusion priors learned from real data to synthetic domains, often sacrificing generalization capabilities for in-the-wild datasets like MVImgNet \cite{yu2023mvimgnet}.

\subsection{Synthetic-to-Real Adaptation} 
Synthetic images provide a rich and diverse source of annotated data, yet models trained on synthetic data often face challenges when generalizing to real-world domains. The Synthetic-to-Real Adaptation task thus aims to leverage synthetic data advantages while ensuring generalization to real-world scenarios. Prior work has addressed this gap by transferring knowledge from synthetic to real domains. For example, knowledge distillation methods \cite{hinton2015distilling} have been applied to semantic segmentation and object detection in LiDAR point clouds \cite{sautier2022image, li2022self, zhang2023pointdistiller}, while transfer learning has improved the real-world performance of autonomous driving models initially trained in simulated environments \cite{akhauri2020enhanced, kim2017end}.
Other researchers have focused on specialized training strategies to improve model generalization by leveraging knowledge from related tasks. For example, inspired by human learning processes, Curriculum Learning \cite{kumar2010self} structures a model's learning path progressively from easy to difficult tasks, which has been applied to detectors trained in simulators, transitioning from simple to complex tasks to achieve robust real-world performance \cite{soviany2021curriculum}. Additionally, reinforcement learning has been used to dynamically generate curriculum strategies, expediting training and enabling models to tackle a broader range of real-world challenges \cite{qiao2018automatically, song2021autonomous, bae2021curriculum, anzalone2022end}.

Transferring models to new tasks often requires retraining for each specific task and model, a process that is both resource-intensive and time-consuming. To address this challenge, many studies have focused on direct synthetic-to-real-domain transfer. Early approaches primarily used unsupervised learning combined with task-specific losses to preserve image content \cite{mueller2018ganerated, murez2018image, zheng2018t2net}. To overcome the limitations of low-level loss and semantic misalignment from feature space transfer, works such as \cite{hoffman2018cycada, imbusch2022synthetic, bousmalis2017unsupervised} focused on aligning data directly in the image space by leveraging GANs' distribution-matching capabilities. \cite{bi2019deep} was among the first to propose disentangling synthetic images into shading and albedo layers and transferring each separately, thus avoiding ambiguities and misalignments between illumination and texture. More recently, diffusion models have demonstrated great success in generation tasks; \cite{azuma2024zodi, li2024aldm} used these models to reframe synthetic-to-real transfer as an image-to-image generation task.

Previous approaches often require training numerous additional modules and employing complex strategies. 
In contrast, we propose that the core challenge of synthetic-to-real image transfer lies in handling unlabeled and unpaired data. Additionally, leveraging a strong diffusion prior requires regularizing the model to consistently produce reliable, transferred images. To address this, we simplify the process by iteratively generating high-quality, reliable pseudo-labels (an approach demonstrated effective in unsupervised domain adaptation \cite{cuttano2023cross, lee2023camera, litrico2023guiding}) to fine-tune a diffusion model, namely IntrinsicAnything \cite{chen2024intrinsicanything}.

\section{Method}\label{sec:method}

\begin{figure*}[tb]
\begin{center}
\includegraphics[width=1\linewidth]{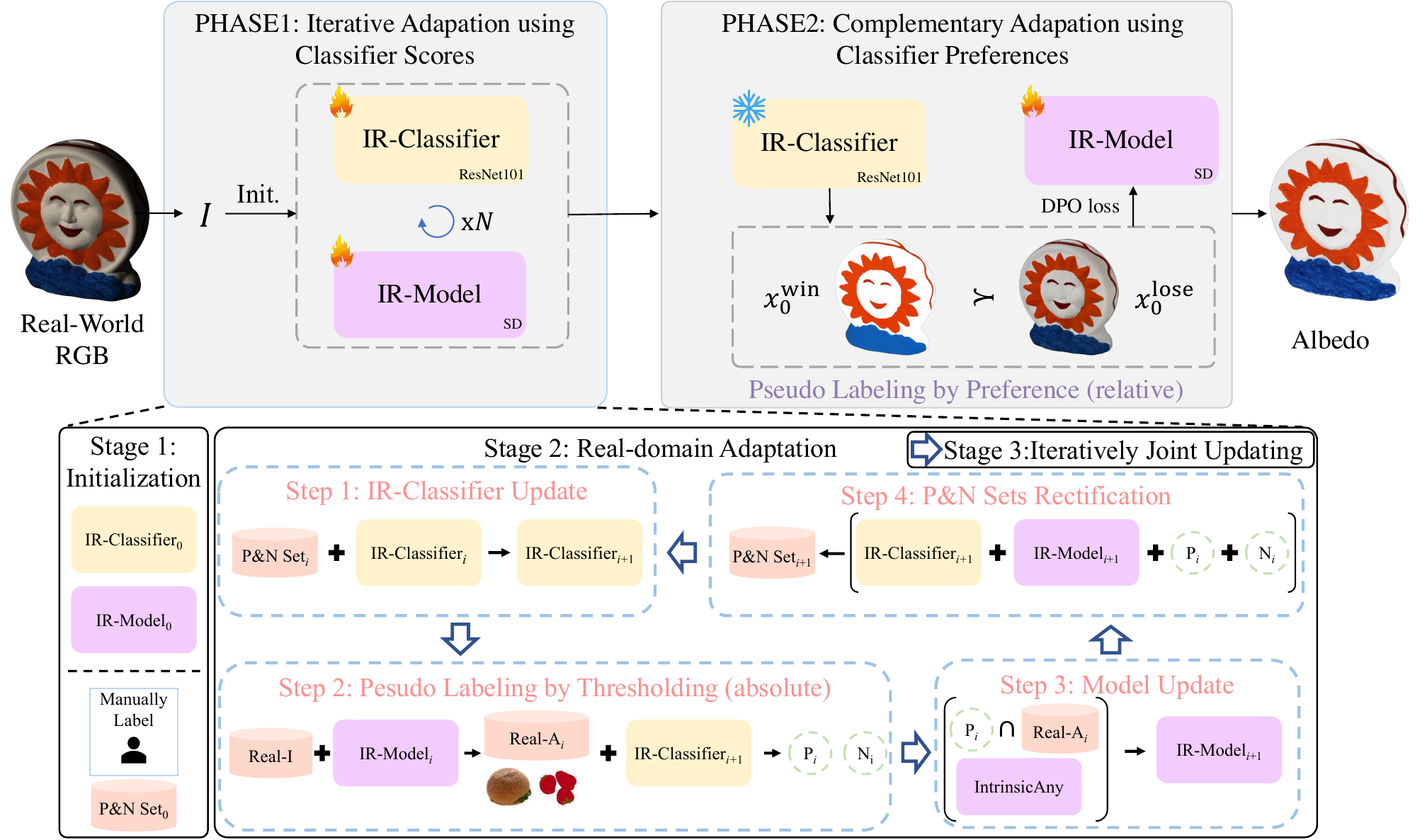}
\end{center}
\caption{
\textbf{Overview of our IntrinsicRreal framework.}
Our IR (IntrinsicReal) framework consists of two phases. i) Pseudo-labeling with an {\bf absolute} confidence threshold on classifier predictions. Specifically, in Stage 1, we initialize the IR-Classifier$_0$, IR-Model$_0$ and P\&N Set$_0$.
Please note that P\&N Set$_0$ is initialized using a small number of manual labels.
In Stage 2, the framework updates the IR-Classifier$_i$, IR-Model$_i$, and P\&N Set$_i$ using Real-I (from the MVImgNet dataset), respectively.
In Stage 3, the iterative joint update strategy gradually improves the performance of the models. Real-A$_i$ refers to the albedo image inferred from IR-Model$_i$ and Real-I. P$_i$ and N$_i$ represent positive and negative pseudo-labels, respectively.
The symbol $\cap$ denotes the use of pseudo-labels (e.g., P$_i$) to select corresponding albedos from Real-A$_i$ for updating IntrinsicAnything.
ii) Pseudo-labeling using the {\bf relative} preference ranking of classifier predictions for individual input objects.
We construct pairs of generated albedos for the same input object $I$ by comparing the output albedos from the last-iteration IR-Model (denoted as $x_0^{win}$) with those from iteration-0 and iteration-1 IR-Models (denoted as $x_0^{lose}$). And then fine-tune the last-iteration IR-Model using the Diffusion-DPO loss.
}
\label{fig:overview}
\end{figure*}

\subsection{Overview} 

\begin{figure}
    \centering
    \vspace{-0.2cm}
    \includegraphics[width=0.92\linewidth]{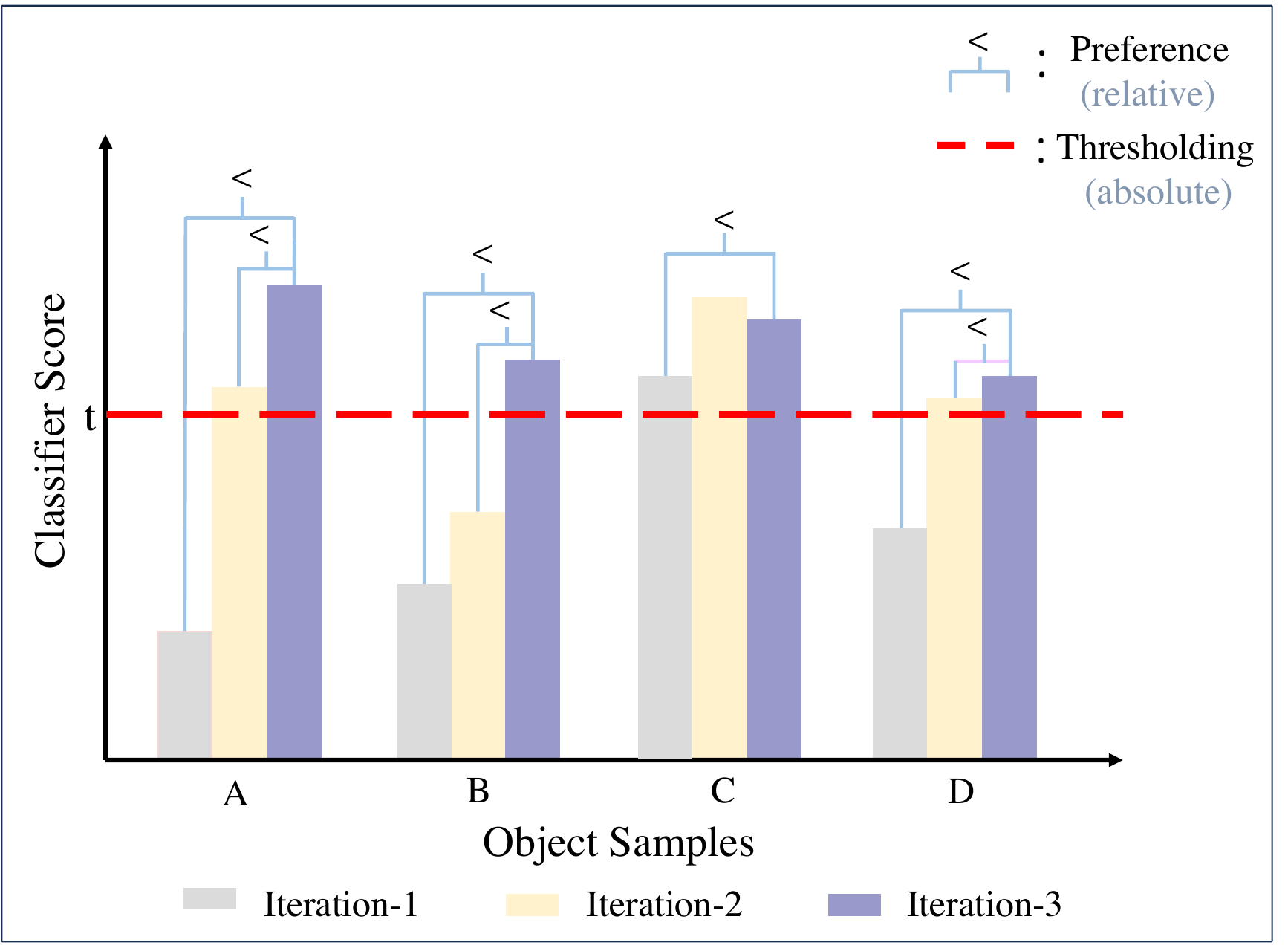}
    \vspace{-0.3cm}
    \caption{\textbf{Illustration of our dual pseudo labeling strategy.} }
    \label{fig:illustration_figs}
    \vspace{-0.5cm}
\end{figure}

\noindent
Our IntrinsicReal aims to enhance the generalization capability of a state-of-the-art intrinsic image decomposition method, \ie, IntrinsicAnything~\cite{chen2024intrinsicanything}, for real-world scenarios.
Specifically, we base our approach on the principle that an intrinsic image decomposition model can achieve robust real-world performance if trained with fully labeled data in a supervised manner. 
However, given that we only have unlabeled RGB data from the MVImgNet dataset, we introduce a novel dual pseudo-labeling strategy that fine-tunes the albedo generation model using only a small amount of human labels.
Specifically, we create two types of pseudo-labels using i) an {\bf absolute} confidence threshold applied to the classifier predictions and ii) {\bf relative} preference rankings derived from different classifier predictions for individual input objects. The illustration of our dual pseudo labeling strategy as shown in Fig.~\ref{fig:illustration_figs}.

To leverage these pseudo-labels, we introduce a novel two-phase pipeline: i) iterative adaptation using classifier scores (Sec.~\ref{sec:iterative_adaptation}) and ii) complementary adaptation using classifier preferences (Sec.~\ref{sec:dpo}), which sequentially apply these labeling techniques.
Please refer to Fig.~\ref{fig:overview} for an overview of our IntrinsicReal.

\subsection{Iterative Adapation using Classifier Scores}
\label{sec:iterative_adaptation}

We introduce a novel iterative joint updating scheme involving two components: i) IntrinsicReal-Model for albedo image generation and ii) IntrinsicReal-Classifier for pseudo-labeling, which are refined iteratively using only a small amount of labeled data.
Note that this pseudo-labeling is achieved by applying an absolute confidence threshold to the classifier outputs.
This scheme comprises three key stages: i) Initialization (Sec.\ref{sec:init_class}); ii) Real-Domain Adaptation (Sec.\ref{sec:real_adapt}); and iii) Iterative Joint Updating (Sec.~\ref{sec:iter_update}).

\subsubsection{ Initialization}\label{sec:init_class}

\noindent \textbf{Initialization of IntrinsicReal-Model.}
We initialize it with a synthetic-trained IntrinsicAnything~\cite{chen2024intrinsicanything} model, and denote it as IntrinsicReal-Model$_0$ (IR-Model$_0$).

\begin{figure}
    \centering
    \vspace{-0.3cm}
    \includegraphics[width=0.9\linewidth]{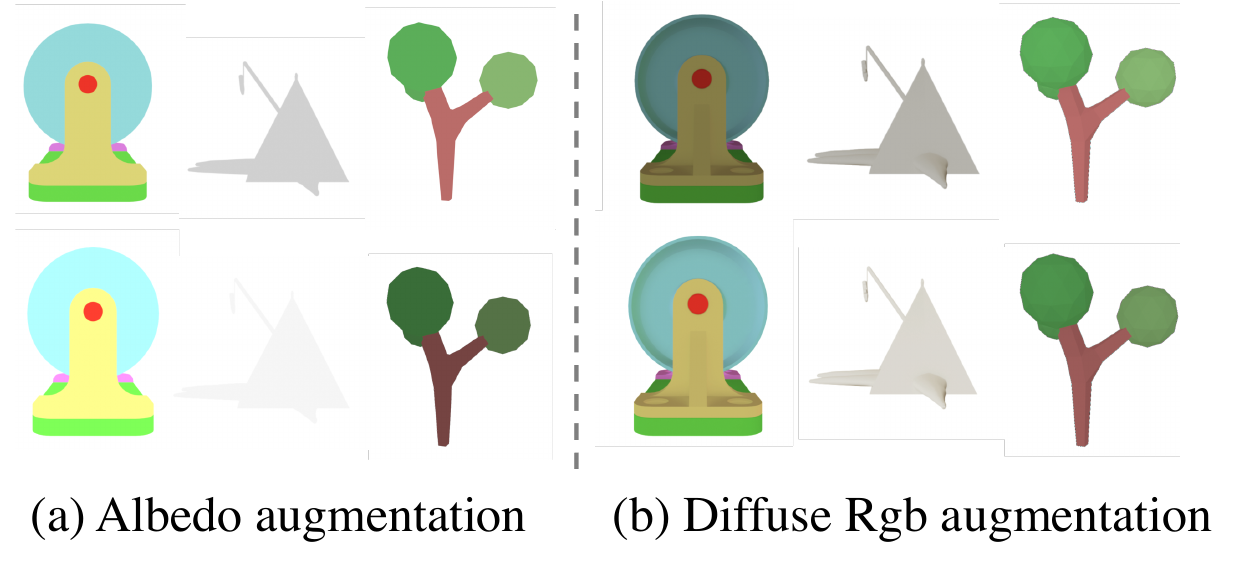}
    \vspace{-0.3cm}
    \caption{\textbf{Visualization of illuminance-aware global augmentation applied to albedo and diffuse rgb.} Top row: raw albedo and diffuse rgb. Bottom row: augmented albedo and diffuse rgb.}
    \label{fig:augmentation}
    \vspace{-0.4cm}
\end{figure}

\vspace{1mm}
\noindent \textbf{Initialization of IntrinsicReal-Classifier.}
To overcome the lack of ground truth albedo data, we propose initializing our IntrinsicReal-Classifier$_0$ (IR-Classifier$_0$) by training an albedo-diffuse rgb classifier on synthetic data:
\begin{itemize}
    \item \textit{Synthetic Data Creation.} 
    We use the Blender Cycles engine to render paired albedo and shading images based on Objaverse~\cite{objaverse}, a recent large-scale dataset of 3D objects within the synthetic domain.
    Specifically, Objaverse provides illumination-invariant albedo image $A(\mathbf{I})$ and illumination-varying shading image $S(\mathbf{I})$ of an object $\mathbf{I}$. Following the Lambertian assumption commonly applied in intrinsic image decomposition, we obtain the corresponding diffuse rgb image $I_{diff}(\mathbf{I})$ using:
    \begin{equation}
        I_{diff}(\mathbf{I}) = A(\mathbf{I}) \odot S(\mathbf{I})
    \label{eq:img_decom}
    \end{equation}
    where $\odot$ represents channel-wise multiplication. 
    Note that we apply illuminance-aware global augmentation to both albedo and diffuse rgb images utilize adjusting image intensity, as shown in Fig.~\ref{fig:augmentation}. Since our adaptation process progresses from easy to difficult, we initially removed the specular variable to constrain the initial classifier's objective, focusing solely on albedo.
    
    \item \textit{Model Architecture}. Given the inherent difficulty in distinguishing albedo from diffuse rgb due to their color similarities, our goal is to design a classifier with sufficient discriminative power without being overly sensitive to easily identifiable features. For example, classifiers based on large pre-trained models like DINO~\cite{caron2021emerging} or CLIP~\cite{radford2021learning} might over-rely on their extensive prior knowledge and focus on albedo-irrelevant features. 
    Therefore, we choose a ResNet101~\cite{he2016deep} pre-trained on ImageNet~\cite{deng2009imagenet} as our classifier backbone.
    Notably, the diffuse rgb images obtained from Eq.~\ref{eq:img_decom} are generally darker than albedo in synthetic data, a contrast not observed in real images. To preserve shadow details and mitigate the impact of color brightness on classification, we apply light and shadow enhancement, ensuring the classifier focuses more effectively on albedo.
\end{itemize}
\vspace{1mm}
\noindent \textbf{Initialization of Positive and Negative Sets.} 
To fine-tune our IntrinsicReal-Classifier, we use two sets of albedo images: a Positive set and a Negative set, consisting of positive and negative samples, respectively. 
Our Positive-Set$_0$ and Negative-Set$_0$ are initialized with a small number of manually annotated positive and negative albedo images generated by IR-Model$_0$, respectively.

\subsubsection{Real-Domain Adaptation}\label{sec:real_adapt}

\noindent
Let $i$ be the current iteration, we present the key techniques for adapting our IntrinsicReal-Model and IntrinsicReal-Classifier to real-world data as follows.

\vspace{1mm}
\noindent \textbf{IntrinsicReal-Classifier Update.} 
We use Positive-Set$_i$ and Negative-Set$_i$ to fine-tune IR-Classifier$_i$, resulting in IR-Classifier$_{i+1}$.

\vspace{1mm}
\noindent \textbf{Pesudo-labeling.}
We apply IR-Classifier$_{i+1}$ to the albedos of MVImgNet data (estimated by IR-Model$_i$) and generate pseudo-labels by applying two confidence score thresholds to the predictions. We set the positive threshold to 0.99 for generating positive pseudo-labels P$_i$ and the negative threshold to 0.3 for generating negative pseudo-lables N$_i$.

\vspace{1mm}
\noindent \textbf{IntrinsicReal-Model Update.}
We then use the paired albedo $A$ and RGB images $I$ corresponding to P-$i$ to fine-tune a synthetic-trained IntrinsicAnything model,
producing IR-Model$_{i+1}$.
Note that we consistently start from IntrinsicAnything to avoid biases introduced by duplicate albedo images appearing across multiple rounds of pseudo-labeling. This approach helps the model retain its generalization ability and prevents it from getting stuck in local optima.
Specifically, we first use a pre-trained VAE image encoder $\mathcal{E}$ to extract the conditional signal feature from $I$ and have $\mathcal{E}(I)$. The diffusion process then adds noise to the encoded latent $z = \mathcal{E}(A)$, producing a noisy latent $z_t$, with $t$ uniformly sampled from $\{1, \ldots, T\}$.
We train a network $\epsilon_{\theta}$ to predict the noise added to $z_t$, conditioned on the image encoding $\mathcal{E}(I)$, by minimizing:
\begin{equation}
    L = \mathbb{E}_{\mathcal{E}(A),\mathcal{E}(I),\epsilon, t}\left[\left|\epsilon-\epsilon_{\theta}\left(z_{t}, t,\mathcal{E}(I)\right)\right|_{2}^{2}\right]\label{eq:1}
\end{equation}
where 
$\epsilon\sim\mathcal{N}(0,1)$ is a standard normal distribution.

\vspace{1mm}
\noindent \textbf{Positive and Negative Sets Rectification.} 
We observe notable improvements in the estimated albedo from IR-Model$_{i+1}$, particularly in regions affected by occlusion and shadowing. Therefore, we construct Positive-Set$_{i+1}$ with the pseudo-labels P$_i$ and their corresponding improved albedos estimated by IR-Model$_{i+1}$.
For Negative-Set$_{i+1}$, we first update the albedos corresponding to pseudo-labels N$_i$ with those estimated by IR-Model$_{i+1}$ and apply IR-Classifier$_{i+1}$ to update their confidence scores.
Then, we remove data with confidence scores larger than 0.5 from N$_i$ and obtain N'$_i$.
Finally, we construct Negative-Set$_{i+1}$ as the pseudo-labels N'$_i$ and their corresponding albedos.

\subsubsection{Iteratively Joint Updating}\label{sec:iter_update}

\noindent
As aforementioned, we implement our iterative joint updating strategy for iterations $i = \{0, 1, 2, ...\}$.
In each iteration, we sequentially perform the IntrinsicReal-Classifier Update, Pseudo-labeling, IntrinsicReal-Model Update, and the Positive and Negative Sets Rectification. 
Please refer to Sec.~\ref{sec:Alg_pusedo_code} in the supplement for the method's pseudo-code.

\subsection{Complementary Adapation using Classifier Preferences}\label{sec:dpo}

\begin{figure}
    \centering
    \vspace{-0.2cm}
    \includegraphics[width=0.85\linewidth]{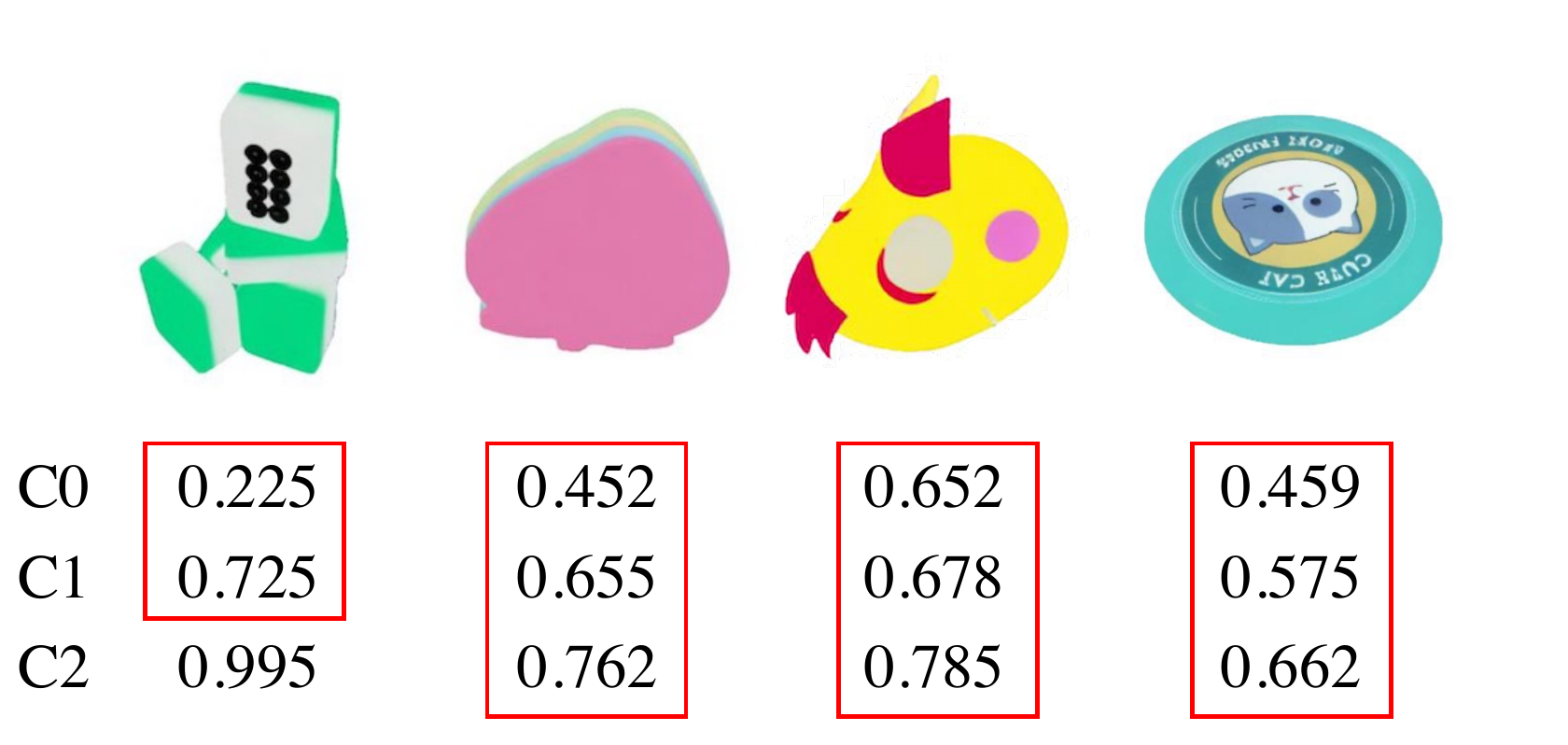}
    \vspace{-0.2cm}
    \caption{\textbf{IR-Classifier score for High-quality albedo.} The threshold is 0.99. Some albedos fall below the threshold despite their high quality.}
    \label{fig:illustration_IR_C}
    \vspace{-0.5cm}
\end{figure}

While effective, Phase 1 is not without limitations, leaving room for improvement. In particular, the pseudo-labeling strategy in Phase 1 relies on an absolute threshold applied to classifier outputs. Although this approach performs well for the highest-quality outputs, it becomes less reliable for sub-optimal cases. As illustrated in Fig.~\ref{fig:illustration_IR_C}, high-quality albedos can receive varying classifier scores, with some falling below the threshold despite their quality.
To address this limitation and leverage these low-score but high-quality outputs during fine-tuning, we propose a complementary pseudo-labeling strategy based on the classifier's relative preference ranking. 
The core idea is that when comparing different generated albedos of the {\it same} input object, the classifier's preference ranking is reliable and can be effectively utilized for fine-tuning.

Specifically, we construct pairs of generated albedos for the same input object image $I$ by comparing the output albedos from the last-iteration IR-Model (denoted as $x_0^w$, which corresponds to iteration-2) with those from iteration-0 and iteration-1 IR-Models (denoted as $x_0^l$). We include a pair only if the classifier output of $x_0^w$ is higher than that of $x_0^l$. 
We then fine-tune the last-iteration IR-Model from Phase 1 using the Diffusion-DPO loss~\cite{wallace2024diffusion} as follows:
\begin{align}
\vspace{-0.8cm}
\mathcal{L}(\theta) = & - \mathbb{E}_{(x_0^w, x_0^l) \sim \mathcal{D}, t \sim \mathcal{U}(0,T), 
x_t^w \sim q(x_t^w | x_0^w, I), x_t^l \sim q(x_t^l | x_0^l, I)} \nonumber \\
& \log \sigma \left( -T \omega (\lambda_t) \right) ( \nonumber \\
& \quad \| \epsilon^w - \epsilon_{\theta} (x_t^w, t, I) \|_2^2 
- \| \epsilon^w - \epsilon_{\text{ref}} (x_t^w, t, I) \|_2^2 \nonumber \\
& \quad - ( \| \epsilon^l - \epsilon_{\theta} (x_t^l, t, I) \|_2^2 
- \| \epsilon^l - \epsilon_{\text{ref}} (x_t^l, t, I) \|_2^2 )) 
\end{align}
\vspace{-0.3cm}
\label{eq:dpo_loss}

\noindent where ``$w$'' represents ``$win$'' and ``$l$'' represents ``$lose$''; $\epsilon^w$ and $\epsilon^l$ denote Gaussian noise for $x_t^w$ and $x_t^l$, respectively; $\epsilon_{ref}$ is Gaussian noise from the pre-trained model, \ie, IR-Model$_{i+1}$;
$\lambda_t$ is the signal-to-noise ratio; $\omega (\lambda_t)$ denote a weighting function. 
Intuitively, $\mathcal{L}$ encourages $\epsilon_\theta$ to be closer to $x_t^w$ and away from $x_t^l$.

\noindent
\section{Experiments}
\label{sec:experiments}

\subsection{Datasets}


\noindent We trained our model on two large-scale datasets: Objaverse \cite{objaverse} and MVImgNet \cite{yu2023mvimgnet}. For evaluation, we employed the MIT Intrinsic Dataset \cite{grosse2009ground} containing 20 real objects with ground-truth intrinsic image decompositions.

\vspace{1mm}
\noindent \textbf{Objaverse (Synthetic) \cite{objaverse}.} The Objaverse dataset~\cite{objaverse} contains over 800K (and growing) 3D models. Given the lack of intrinsic images in Objaverse, we generated paired albedo and irradiance images using the Blender Cycles engine, randomly selecting 30K objects rendered under diverse HDR environment maps. Models with low-quality albedo or irradiance images (\eg, black or white images) were excluded. Additionally, we derived paired diffuse rgb images using Eq.~\ref{eq:img_decom}.


\vspace{1mm}
\noindent \textbf{MVImgNet (Real)~\cite{yu2023mvimgnet}.}
MVImgNet is a large-scale dataset of multi-view images, comprising 6.5 million frames across 238 object classes. Unlike Objaverse, MVImgNet consists entirely of real-world data. To enhance classifier training, we performed a series of preprocessing steps on the MVImgNet RGB data:

\begin{itemize}
    \item \textbf{Data Collection}: 40,000 RGB images from 200 object categories were randomly selected.
    \item \textbf{Albedo Processing}: Estimated albedo maps for all RGB images using IntrinsicAnything, followed by quality filtering to remove non-informative cases (e.g., pure black/white albedo-RGB pairs).
    \item \textbf{Dataset Curation}: Split into 30,000 training and 3,000 validation samples. The validation set underwent rigorous annotation by 7 graphics experts using a tri-class system (positive/negative/ambiguous). 
\end{itemize}

\subsection{Implementation Details}

\noindent \textbf{Implementation of IntrinsicReal-Classifier.} Due to inherent ambiguities between albedo and shading, it is crucial for the classifier to have sufficient capacity in the initial stage to promote efficient convergence while avoiding overly strong priors that could shift classification away from using albedo features. 
Therefore, we select a ResNet101~\cite{he2016deep} pre-trained on ImageNet~\cite{deng2009imagenet} as the backbone for our classifier.
We implemented our IntrinsicReal-Classifier in PyTorch and optimized them using Adam~\cite{kingma2014adam} with a learning rate of $5 \times 10^{-4}$ for 250K iterations.


\vspace{1mm}
\noindent \textbf{Implementation of IntrinsicReal-Model}.
We create it by fine-tuning the IntrinsicAnything model~\cite{chen2024intrinsicanything}, which is based on the Image Condition Stable Diffusion model~\cite{rombach2022high}. 
We train our model for 100K iterations using the Adam optimizer~\cite{kingma2014adam} with a learning rate of $1 \times 10^{-5}$. For the DPO finetuning, we train final iteration model for 10K iterations.


\subsection{Evaluation Details}
\noindent \textbf{Metrics}. We evaluate albedo visual quality using PSNR, SSIM, and MSE metrics between the predicted and ground-truth images in MIT intrinsic dataset~\cite{grosse2009ground} including 20 objects. Additionally, we evaluate the {\it precision} and {\it accuracy} of our classifier to assess its performance at each stage.

\vspace{1mm}
\noindent \textbf{User Study}. 
Considering that the data categories in the MIT dataset are not representative of most real-world scenarios and the dataset size is relatively small, we also conducted evaluations on the validation set of MVImageNet. Since ground-truth albedo is unavailable, we performed a user study to assess the quality of albedo reconstruction. Specifically, we invited 18 professionals, comprising experts and designers in the field of image rendering, to assess and vote on comparative results between our albedo outputs at different iterations and those from IntrinsicAnything. 
Please see the supplementary materials for more details.

\begin{figure}[tb]
\begin{center}
\includegraphics[width=1.0\linewidth]{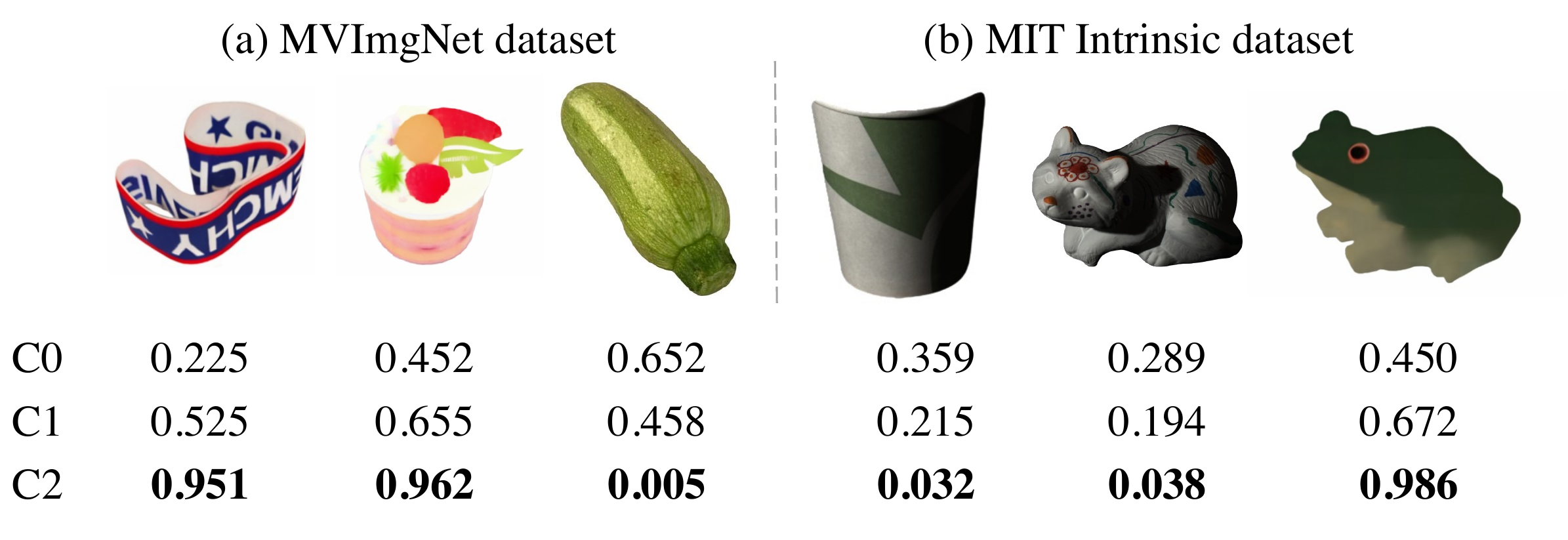}
\end{center}
\vspace{-0.5cm}
\caption{\textbf{Scores of IR-Classifier$_0$~(C0), IR-Classifier$_1$~(C1), and IR-Classifier$_2$~(C2) on the same images.} Score represents the positive labeled albedo confidence value.}
\label{fig:IR_C_ablation}
\vspace{-0.5cm}
\end{figure}

\subsection{Comparison with SOTA Method}
\noindent \textbf{Baselines}. We compare the generalization performance of our method on real-world data against state-of-the-art methods for albedo estimation. Specifically, we compare against IntrinsicAnything~\cite{chen2024intrinsicanything} and RGB-X~\cite{zeng2024rgb}.

\vspace{1mm}
\noindent\textbf{Results}. From the albedo image estimation metrics in Table.~\ref{tab:albedo_compare} and Fig.~\ref{fig:baseline_compare}, we report our the quantitative and qualitative results outperforms IntrinsicAnything~\cite{chen2024intrinsicanything} and RGB-X~\cite{zeng2024rgb} significantly.
Notably, the results obtained from IntrinsicAnything \cite{chen2024intrinsicanything} reveal a significant number of black albedos, particularly for metallic objects, suggesting its limited generalization capabilities for such objects. 
It can also be observed that RGB-X~\cite{zeng2024rgb} demonstrates unsatisfactory performance in real-world objects, such as texture detail degradation and systematic color deviations.
In contrast, our method can significantly reduce the ambiguities related to these objects. Our method exhibits a remarkable capability to handle highlight and shadow conditions, particularly metallic object, and this ability is crucial for achieving precise and reliable mesh with albedo material estimation for downtown tasks.



\begin{table}[htbp]
  \centering
\vspace{-0.3cm}
\resizebox{1.\linewidth}{!}{
  \tiny
  \begin{tabular}{l|ccc}
    \toprule
    & PSNR$\uparrow$  &  SSIM $\uparrow$ & MSE $\downarrow$ 
 \\
    \midrule
    RGB-X~\cite{zeng2024rgb} & 10.709 & 0.429 & 0.117  \\
    IntrinsicAnything~\cite{chen2024intrinsicanything} & 15.765 & 0.731 & 0.033 \\
    \textbf{IntrinsicReal(Ours)} & \textbf{17.449} & \textbf{0.758} & \textbf{0.024} \\
    \bottomrule
  \end{tabular}
}
\vspace{-0.3cm}
\caption{Quantitative results for real-world object albedo estimation. All scores are calculated as an average across 20 objects from the MIT dataset~\cite{grosse2009ground}.}
\label{tab:albedo_compare}
\vspace{-0.1cm}
\end{table}

\begin{figure*}[tb]
\begin{center}
\includegraphics[width=0.92\linewidth]{figures/baseline_compare5.pdf}
\end{center}
\vspace{-6mm}
\caption{
\textbf{Qualitative comparisons with IntrinsicAny~\cite{chen2024intrinsicanything} and RGB-X~\cite{zeng2024rgb} on MIT dataset~\cite{grosse2009ground} and MVImgNet dataset~\cite{yu2023mvimgnet}.}
}
\label{fig:baseline_compare}
\vspace{-0.7mm}
\end{figure*}

\begin{table}[htbp]
  \centering
\resizebox{1\linewidth}{!}{
  \tiny
  \begin{tabular}{l|ccc}
    \toprule
     & \multicolumn{3}{c}{Albedo}\\
    & Acc$\uparrow$  &  Negative Precision$\uparrow$  & Positive Precision$\uparrow$ 
 \\
    \midrule
    IR-Classifier$_0$ & 0.52 & 0.46 & 0.75 \\
    IR-Classifier$_1$ & 0.79 & 0.87 & 0.72 \\
    IR-Classifier$_2$ & \textbf{0.82} & \textbf{0.88} & \textbf{0.78} \\
    \bottomrule
  \end{tabular}
}
\vspace{-0.3cm}
\caption{Quantitative results of ablation study for different iteration IR-Classifier. All scores are calculated across 3,000 objects from the MVImgNet~\cite{yu2023mvimgnet} validation dataset.}
\label{tab:ablation_IR_C}
\vspace{-0.4cm}
\end{table}
\subsection{Ablation Studies}
\noindent
We conduct ablation studies to analyze the contribution of each component in our IntrinsicReal framework using the MVImgNet dataset and MIT dataset, mainly containing Iterative adaptation using Classifier Scores strategy and Complementary Adaptation using Classifier Preferences strategy.
For simplicity, we denote the latter as {\it DPO strategy}.


Specifically, we firstly conduct an ablation study of the IR-Model and IR-Classifier on the validation dataset. The evaluation protocol employs distinct metrics for each component: visual quality metrics for the IR-Model, while the IR-Classifier is measured through comprehensive classification performance indicators including overall accuracy, positive predictive, and negative predictive.

\begin{table}[htbp]
  \centering
\vspace{-0.1cm}
\resizebox{1\linewidth}{!}{
  \tiny
  \begin{tabular}{l|ccc}
    \toprule
    & PSNR$\uparrow$  &  SSIM $\uparrow$  & MSE $\downarrow$ 
 \\
    \midrule
    IR-Model$_0$ & 15.765 & 0.731 & 0.033  \\
    IR-Model$_1$ & 16.309 & 0.732 & 0.031 \\
    IR-Model$_2$ & 16.627 & 0.753 & 0.028 \\
    \textbf{IR-Model$_2$ + DPO} & \textbf{17.449} & \textbf{0.758} & \textbf{0.024} \\
    \bottomrule
  \end{tabular}
}
\vspace{-0.3cm}
\caption{Qualitative results of ablation study for different iterations IR-Model. All scores are calculated as an average across 20 objects from
the MIT dataset.}
\label{tab:ablation_IR_M}
\vspace{-0.1cm}
\end{table}

\begin{figure}[tb]
\begin{center}
\vspace{-0.3cm}
\includegraphics[width=0.93\linewidth]{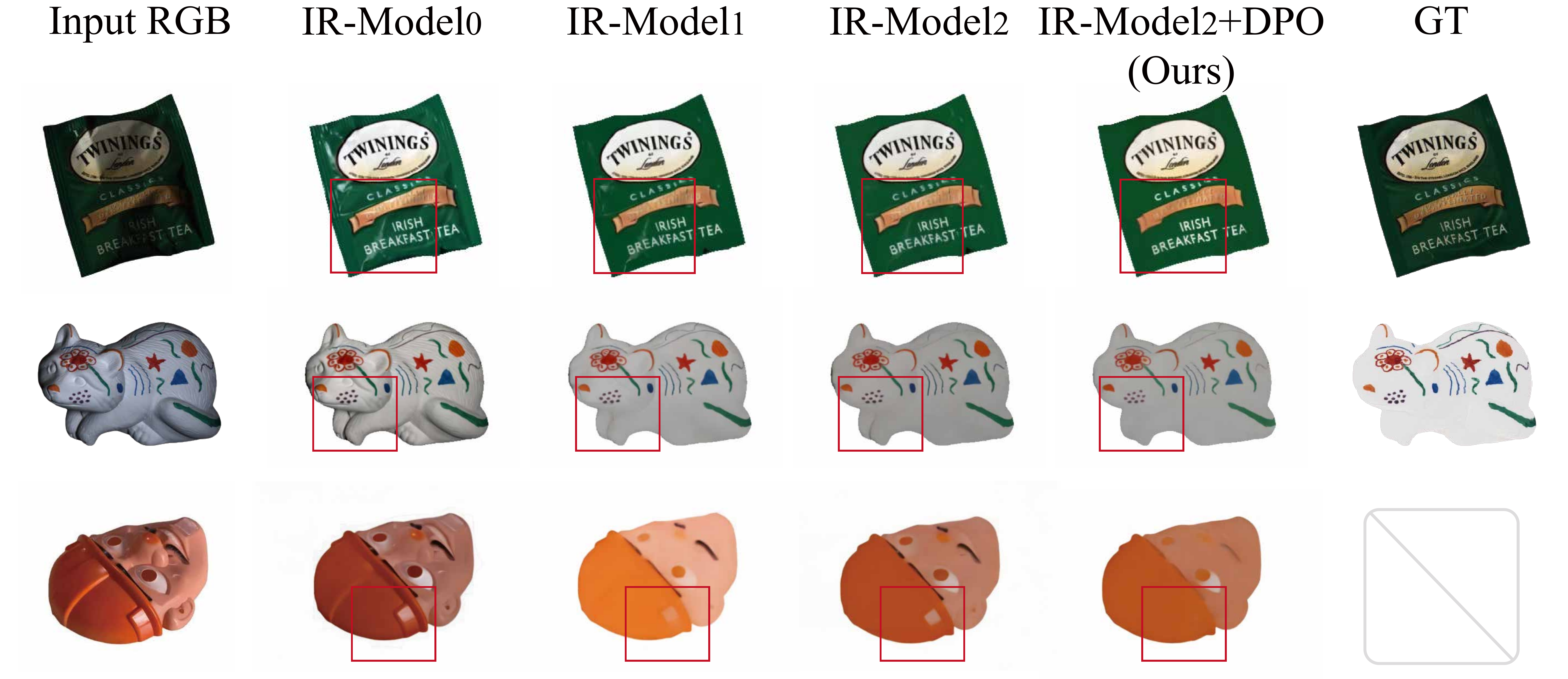}
\end{center}
\vspace{-0.6cm}
\caption{Qualitative results of ablation study for different iterations IR-Model on the MIT dataset and MVImgNet dataset.}
\vspace{-0.2cm}
\label{fig:IR_Model_ablation}
\end{figure}

\noindent \textbf{IR-Model.}
As shown in Table.~\ref{tab:ablation_IR_M} and Fig.~\ref{fig:IR_Model_ablation}, we conduct ablation studies to analyze the contribution of iteratively IR-Model and DPO strategy. The results show steady improvement in the IR-Model's performance. Adding DPO leverages the classifier's reliable preference ranking to enhance fine-tuning, mitigating its absolute error and intrinsic limitations from IntrinsicAny~\cite{chen2024intrinsicanything}.

\vspace{0.5mm}
\noindent \textbf{IR-Classifier.}
Moreover, we perform an ablation study on each iteration of the IR-Classifier, and the results indicate that the IR-Classifier progressively enhances the reliability and accuracy of the pseudo-labels, leading to more robust model performance.
The quantitative and qualitative results as shown in Table.~\ref{tab:ablation_IR_C} and Fig.~\ref{fig:IR_C_ablation}, respectively.

\begin{table}[htbp]
  \centering
\vspace{-0.1cm}
\resizebox{1\linewidth}{!}{
  \tiny
  \begin{tabular}{l|c}
    \toprule
    & Negative Class Ratio$\downarrow$
 \\
    \midrule
    IR-Model$_1$ / IR-Model$_2$ & 0.323 \\
    IR-Model$_1$ + DPO / IR-Model$_2$ & 0.151 \\
    IR-Model$_1$ + DPO / IR-Model$_2$ + DPO & 0.085 \\
    IR-Model$_1$ / IR-Model$_2$ + DPO & \textbf{0.081} \\
    \bottomrule
  \end{tabular}
}
\vspace{-0.3cm}
\caption{User Study results of ablation study for different iterations introducing DPO. All scores are calculated across 500 objects from the MVImgNet~\cite{yu2023mvimgnet} validation dataset.}
\label{tab:ablation_IR_DPO}
\vspace{-0.2cm}
\end{table}

\noindent \textbf{DPO Usage.}
Additionally, we conduct an ablation study to evaluate the impact of the DPO fine-tuning strategy across different iterations of the IR-Model, with the results presented in Table~\ref{tab:ablation_IR_DPO}. 
Given the MIT dataset contains only 20 objects, its limited scale results in insufficient occurrence of negative-labeled cases to statistically validate model improvements. 
We therefore adopt {\it User Study} to evaluate the negative cases ratio in 500 randomly samples on the MVImgNet dataset.
Experiments show comparable results between applying DPO to both IR-Models and using DPO only on IR-Model$_2$. The consistent distribution of negative-labeled samples and stable patterns of loss samples in DPO's win-lose pairs result in similar negative case ratios. For efficiency, we adopt to add DPO on IR-Model$_2$.

\subsection{Limitations and future work}

\noindent \textbf{Limitations.} 
A small proportion of objects may still degrade over iterations due to ambiguities between albedo, lighting, and shadows. While our method partially mitigates issues with mirror-like surfaces and persistent shadows, achieving perfect albedo recovery remains unsolved.
\noindent \textbf{Future Work.}
Given the scarcity of RGB-albedo paired datasets and the critical role of albedo in physically-based rendering (PBR), we plan to leverage our classifier for large-scale annotation, advancing scaling law applications.

\section{Conclusion}\label{sec:conclusion}

\noindent
In this work, we propose IntrinsicReal, a novel domain adaptation framework that bridges the domain gap between synthetic and real-world data for intrinsic image decomposition. Our approach adapts IntrinsicAnything to real-world images by fine-tuning it with high-quality albedo outputs, selected through a dual pseudo-labeling strategy. This strategy combines absolute confidence thresholds and relative preference rankings, inspired by human evaluation processes.
By introducing a two-phase adaptation pipeline, IntrinsicReal effectively refines albedo estimation in real-world scenarios, improving its generalization beyond synthetic datasets.

\small
\bibliographystyle{ieeenat_fullname}
\bibliography{main.bib}

\clearpage

\appendix
\section*{Supplementary Materials}



\section{Details of Methods}

\subsection{Classifier Training}
As mentioned in Sec.~2 of the main paper, we use a pretrained ResNet101~\cite{he2016deep} network to implement our binary classifier, and we utilize cross-entropy loss to optimize our network, the details are as follows:
\begin{equation}
    L = \frac{1}{N} \sum_{i=1}^{N} L_i = -\frac{1}{N} \sum_{i=1}^{N} [y_i \log(\hat{p}_{i1}) + (1-y_i) \log(1-\hat{p}_{i1})],
\end{equation}
where $N$ is the number of samples in the dataset, $L$ is the average cross-entropy loss value over the entire dataset, $y_i$ is the true label of the $i$-th sample, taking values 0 or 1, corresponding to our negative and positive classifications, respectively. $\hat{p}_{i1}$ is the predicted probability that the $i$-th sample belongs to the positive class.

\subsection{Iteratively Joint Updating Algorithm}\label{sec:Alg_pusedo_code}

\noindent
As aforementioned, we implement our iterative joint updating strategy for iterations $i = \{0, 1, 2, ...\}$.
In each iteration, we sequentially perform the IntrinsicReal-Classifier Update, Pseudo-labeling, IntrinsicReal-Model Update, and the Positive and Negative Sets Rectification.
Please see Alg.~\ref{alg:method} for the pseudo-code of our method.

\begin{algorithm}
\caption{Simplified Iterative Model Updating and Rectification Process}
\begin{algorithmic}[1]
\State \textbf{Initialization}
    \State Initialize classifier $C_0$
    \State Initialize model $M_0$
    \State Initialize positive and negative sets $P^{\rm set}_0\&N^{\rm set}_0$
 
\For{$i = 1$ to $n$}
    \State \textbf{Update Classifier}
        \State $C_i \leftarrow \text{TrainClassifier}(C_{i-1}, P^{\rm set}_{i-1}\&N^{\rm set}_{i-1})$
    
    \State \textbf{Pseudo Labeling}
        \State $P_i\&N_i \leftarrow \text{PseudoLabel}(C_i, \text{UnlabeledData})$
     
    \State \textbf{Update Model}
        \State $M_i \leftarrow \text{TrainModel}(M_{i-1}, P_i\&N_i, \text{LabeledData})$ 
    
    \State \textbf{Rectification of P\&N Sets}
        \State $P^{\rm set}_i\&N^{\rm set}_i \leftarrow \text{RectifySets}(M_i, P_i\&N_i)$
 
\EndFor
 
\State \textbf{Output}
    \State Final classifier $C_n$
    \State Final model $M_n$
    \State Final positive and negative sets $P^{\rm set}_n\&N^{\rm set}_n$
\end{algorithmic}
\label{alg:method}
\end{algorithm}

\section{Additional Experimental Results}

\begin{figure}[tb]
\begin{center}
\includegraphics[width=1.0\linewidth]{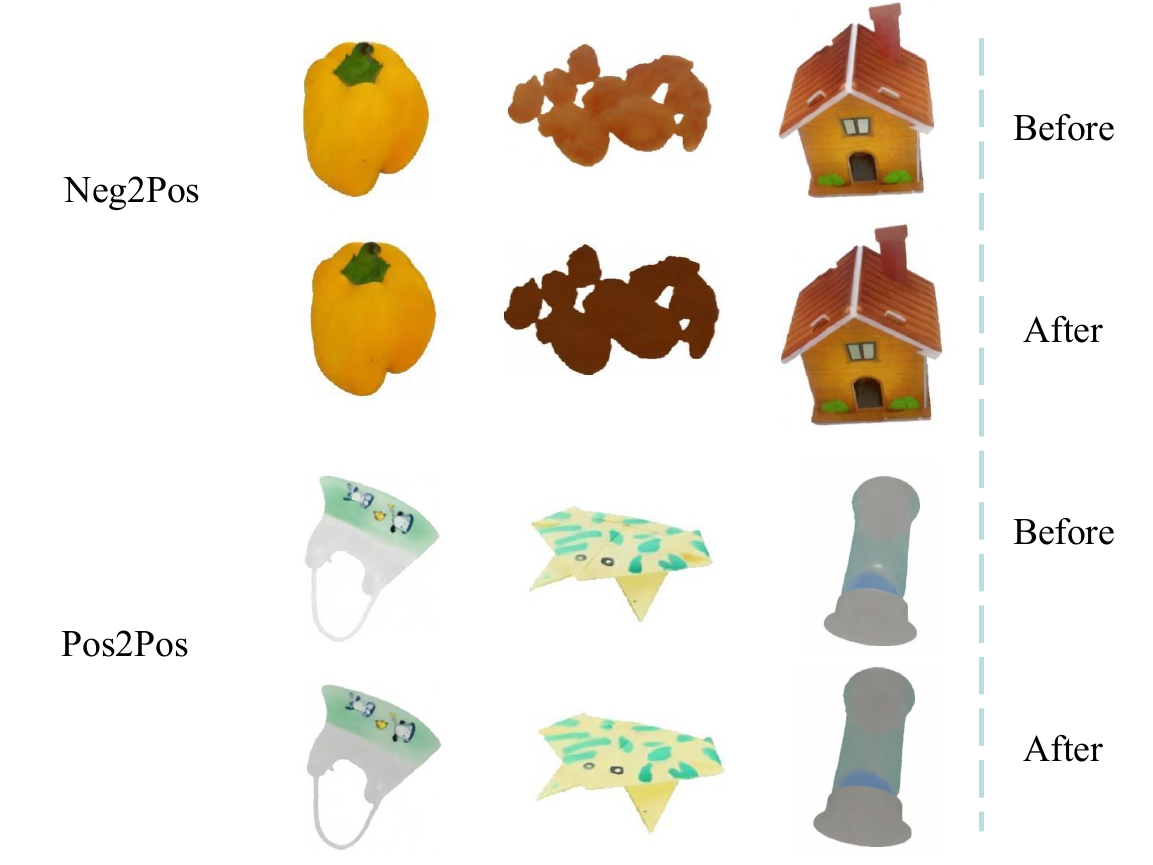}
\end{center}
\vspace{-4mm}
\caption{
\textbf{Visual Comparison of Rectification Step inferred from IR-Model$_2$. } 
Neg2Pos samples are easily reclassified from negative to positive, while Pos2Pos further enhances the quality of the albedo for positive cases.}
\label{fig:rect_compare}
\vspace{-0.5cm}
\end{figure}

\subsection{Robustness Evaluation of the DPO Strategy}
We apply DPO strategy by leveraging the classifier's reliable preference ranking to enhance fine-tuning, mitigating its absolute error and intrinsic limitations from IntrinsicAny~\cite{chen2024intrinsicanything}. To account for the potential presence of suboptimal cases within the "win" class, we conducted additional robustness experiments by introducing 5\% noise into the "win" samples. The results demonstrate that our method maintains consistent performance, confirming its robustness, as shown in Table.~\ref{tab:robustness_IR_M}.

\begin{table}[htbp]
  \centering
\vspace{-0.1cm}
\resizebox{1\linewidth}{!}{
  \tiny
  \begin{tabular}{l|ccc}
    \toprule
    & PSNR$\uparrow$  &  SSIM $\uparrow$  & MSE $\downarrow$ 
 \\
    \midrule
    \textbf{IR-Model$_2$ + DPO} & 17.449 & 0.758 & 0.024 \\
    \textbf{IR-Model$_2$ + DPO + Noise} & 17.401 & 0.743 & 0.022 \\
    \bottomrule
  \end{tabular}
}
\vspace{-0.3cm}
\caption{Qualitative results of robustness validation on the MIT dataset for IR-Model.}
\label{tab:robustness_IR_M}
\vspace{-0.4cm}
\end{table}

\subsection{Details of Rectification Step}
We compare the rectification results in step 4 of real-domain adaptation, as shown in Fig.~\ref{fig:rect_compare}. Specifically, we use IR-Model$_2$ to convert negative cases into positive ones. For example, the first row in the figure presents cases that are ambiguous and can be easily reclassified from negative to positive. In subsequent classifiers, we remove these cases from the Negative-Set 
N, thereby enhancing the reliability of the negative pseudo-labels. Moreover, as illustrated in the second row of Fig.~\ref{fig:rect_compare}, IR-Model$_2$ further improves the quality of the albedo for positive cases.

\subsection{Results of IntrinsicAnything on MVImgNet Dataset}
 We apply IntrinsicAnything~\cite{chen2024intrinsicanything} on the MVImgNet dataset (Fig.~\ref{fig:fig2_supp}), and observe that it exhibits some generalization capability on real-world images. However, it still struggles to accurately estimate the albedo for most objects. Therefore, we can improve the generalization of IntrinsicAnything in real-world scenarios based on this method.

\subsection{Results of Comparison on Synthetic Dataset}
Although we successfully adapt IntrinsicAnything~\cite{chen2024intrinsicanything} from the synthetic to the real object domain, we do not sacrifice performance on synthetic data. In fact, our approach leads to improved results on certain synthetic datasets. By leveraging our strategy, the model achieves enhanced performance across both synthetic and real object domains, as shown in Fig.~\ref{fig:compare_syn}.

\subsection{Results of Comparison on MVImgNet Dataset and MIT Dataset}
\begin{figure*}[tb]
\begin{center}
\includegraphics[width=1\linewidth]{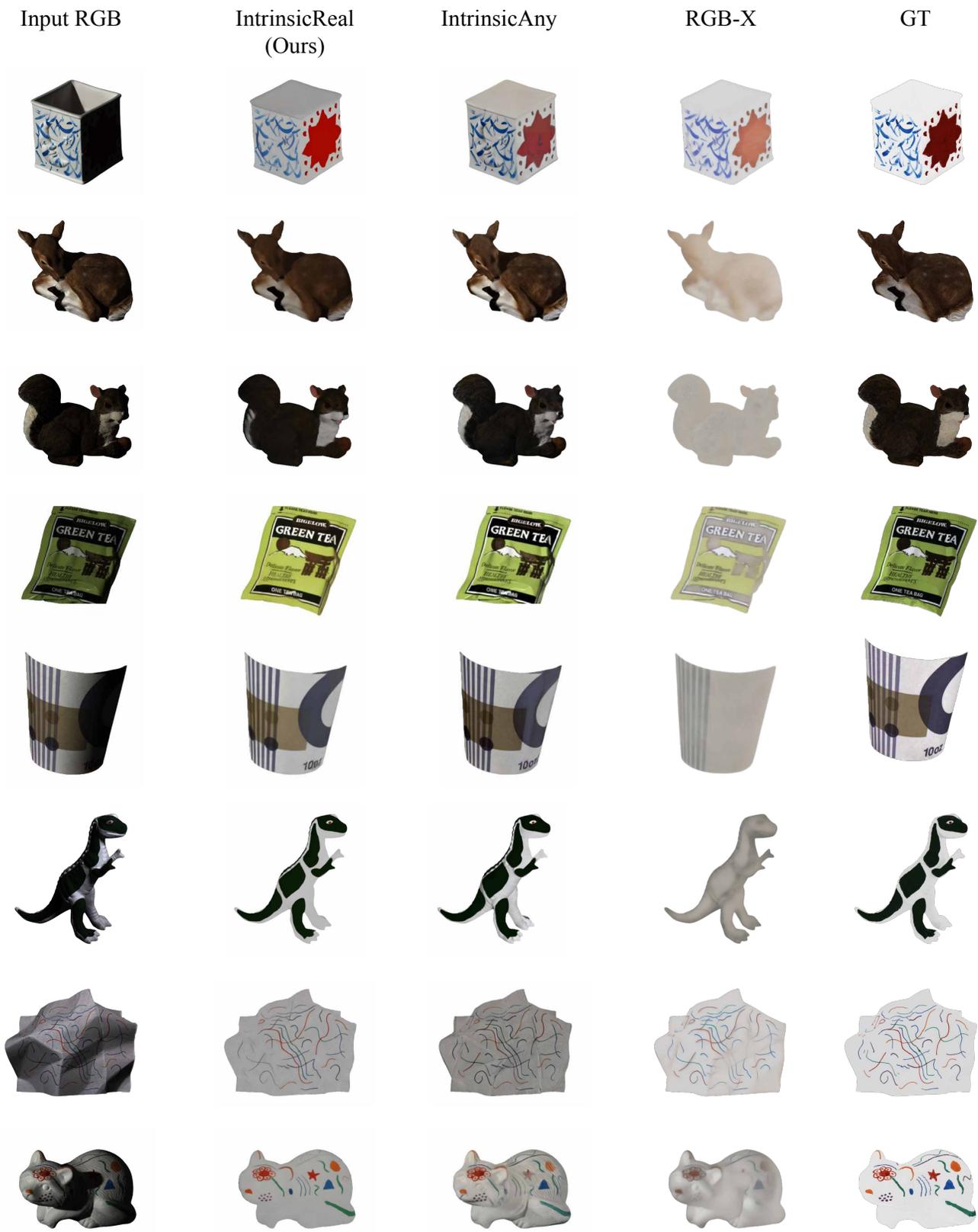}
\end{center}
\caption{
\textbf{Qualitative comparisons with IntrinsicAny~\cite{chen2024intrinsicanything} and RGB-X~\cite{zeng2024rgb} on MIT dataset~\cite{grosse2009ground}.}
}
\label{fig:supp_compare_MIT}
\vspace{-0.3cm}
\end{figure*}

We present additional results to evaluate the generalization performance of our method on the MVImgNet dataset, benchmarked against IntrinsicAnything~\cite{chen2024intrinsicanything} and RGB-X~\cite{zeng2024rgb}. The results are shown in Fig.~\ref{fig:supp_compare_MIT}, Fig.~\ref{fig:supp_compare_MVImgeNet1}, Fig.~\ref{fig:supp_compare_MVImgeNet2}, and Fig.~\ref{fig:supp_compare_MVImgeNet3}.

\subsection{Results of Comparison on Different Iterations of IR-Classifier}
In Fig.~\ref{fig:sup_c1c2_2} and Fig.~\ref{fig:sup_c1c2_1}, we provide additional results that highlight the effectiveness of our IR-Classifier at different stages. Specifically, these results demonstrate that IR-Classifier 1 achieves higher accuracy than IR-Classifier 2 for the same object. By iteratively updating our IntrinsicReal-Classifier, we significantly improve the accuracy of our classifier, enhancing its classification capabilities for both positive and negative cases.

\subsection{Results of Comparison on Different Iterations of IR-Model}
In Fig.~\ref{fig:fig7_supp} and Fig.~\ref{fig:fig7_supp2}, we provide additional results that highlight the effectiveness of our IR-Classifier at different stages. Specifically, these results demonstrate that IR-Classifier 1 achieves higher accuracy than IR-Classifier 2 for the same object. By iteratively updating our IntrinsicReal-Classifier, we significantly improve the accuracy of our classifier, enhancing its classification capabilities for both positive and negative cases.

\subsection{Detailed Ablation Studies for Iterative Joint Update Strategy}

\noindent
As shown in Table~\ref{tab:ablation_study}, we ablate each component of IntrinsicReal on the MVImgNet dataset, evaluating performance with overall accuracy, positive precision, and negative precision.

\begin{itemize}

\item \textit{Stage1: Initialization.}
\begin{itemize}
    \vspace{1mm}
    \item We first perform an ablation study on the IR-Classifier$_0$ with and without the augmentation strategy using synthetic data from Objaverse. 
    Specifically, we ablate the illuminance-aware global augmentation applied to albedo and shading images: $A_0$ / $S_0$ ({\bf A}) and $A = A_0 + {\rm aug}$ / $S = S_0 + {\rm aug}$ ({\bf A1}). 
    As shown in Table~\ref{tab:ablation_study}, applying illuminance-aware data augmentation improves the performance of our classifier.
    \vspace{1mm}
    \item We Then conduct an ablation study on the initialization of Positive-Set$_0$ and Negative-Set$_0$ using manually annotated data. 
    Since they are used to fine-tune IR-Classifier$_0$, we ablate the use of using i) only manually annotated data ({\bf B}) and ii) both manually annotated data and the synthetic data from Objaverse ({\bf B1}). 
    The results indicate that while {\bf B1} shows improvement over IR-Classifier$_0$, it still performs worse than {\bf B}.
    We attribute this gap to the interference caused by the domain gap between synthetic and real data.
    \vspace{1mm}
    \item We also investigate an alternative strategy where synthetic albedos from Objaverse are used as initial Positive set samples, and MVImgNet albedos estimated by IntrinsicAnything are used as initial Negative set samples. (\textbf{B2}). 
    The rationale behind this approach is that synthetic data provides ground-truth albedo images, while IntrinsicAnything's estimated albedos may be less accurate for real-world data. 
    Experimental results reveal that although this strategy improves positive class precision, it also leads to a significant number of misclassifications, where negative samples are wrongly classified as positive. 
    This outcome further supports the efficacy of using manually annotated data for the Positive and Negative sets initialization. Fig.~\ref{fig:compare_p_n} shows some examples of positive and negative results.
    
\end{itemize}

\vspace{1mm}
\item \textit{Stage 2: Real-Domain Adaptation}.
\begin{itemize}
    \vspace{1mm}
    \item To justify the effectiveness of the pseudo-labels generated by our IntrinsicReal-Classifier, we fine-tune IR-Classifier-1 with the pseudo-labels it generated ({\bf C}). As shown in Table~\ref{tab:ablation_study}, all classifier metrics demonstrate a significant improvement, underscoring the effectiveness of our pseudo-labels.  Building on this, we further refine the Positive and Negative sets by enhancing the quality of albedo and eliminating ambiguous instances that exhibit uncertainty. This iterative refinement process aims to progressively boost the reliability and accuracy of the pseudo-labels, leading to more robust model performance.
    \vspace{1mm}
    \item To justify the effectiveness of our Positive and Negative set rectification strategy, we ablate it as ({\bf C}, without rectification) and ({\bf D}, with rectification). The results show that our rectification significantly enhances the classifier's accuracy, leading to a further improvement in positive precision. The qualitative results as shown in Fig.~\ref{fig:c1c2_score}
\end{itemize}

\vspace{1mm}
\item \textit{Stage 3:Iteratively Joint Updating}.
We iterate Stage 2 to further refine our Intrinsic-Model, Intrinsic-Classifier, and Positive and Negative sets (\textbf{E}), which show consistent improvements in both accuracy and precision, validating our initial hypothesis. 

\end{itemize}
\begin{table}[t]  
\centering
\caption{Ablation Studies. All scores are calculated across 3,000 objects from the MVImgNet~\cite{yu2023mvimgnet} validation dataset.}
\vspace{-0.3cm}
\resizebox{0.95\linewidth}{!}{
  \begin{tabular}{l|ccc}
    \toprule
     & \multicolumn{3}{c}{Albedo}\\
    & Acc$\uparrow$  &  Negative Precision$\uparrow$  & Positive Precision$\uparrow$ 
 \\
    \midrule
    A0 & 0.50 & 0.41 & 0.56 \\
    A & 0.52 & 0.46 & 0.75 \\
    B1 & 0.71 & 0.72 & 0.70 \\
    B2 & 0.52 & 0.43 & \textbf{0.86} \\
    B & 0.79 & 0.87 & 0.72 \\
    C & 0.81 & \textbf{0.90} & 0.75 \\
    D & 0.82 & 0.88 & 0.78 \\
    E & \textbf{0.84} & \textbf{0.89} & \textbf{0.79} \\
    \bottomrule
  \end{tabular}
}
\label{tab:ablation_study}
\vspace{-0.4cm}
\end{table}

\section{Applications}
After successfully decomposing intrinsic properties from in-the-wild images, our \textbf{IntrinsicReal} method enables highly realistic object editing by manipulating these properties. As demonstrated in Fig.~\ref{fig:relighting}, our approach achieves strong relighting results on both in-the-wild and synthetic data.

\begin{figure*}[tb]
\begin{center}
\includegraphics[width=0.82\linewidth]{supp_figs/compare_MVimgNet1.pdf}
\end{center}
\vspace{-6mm}
\caption{
\textbf{Qualitative comparisons with IntrinsicAny~\cite{chen2024intrinsicanything} and RGB-X~\cite{zeng2024rgb} on MVImgNet dataset~\cite{yu2023mvimgnet}.}
}
\label{fig:supp_compare_MVImgeNet1}
\vspace{-0.3cm}
\end{figure*}
\begin{figure*}[tb]
\centering
\includegraphics[width=0.78\linewidth]{supp_figs/compare_MVimgNet2.pdf}
\vspace{-0.6cm}
\caption{
\textbf{Qualitative comparisons with IntrinsicAny~\cite{chen2024intrinsicanything} and RGB-X~\cite{zeng2024rgb} on MVImgNet dataset~\cite{yu2023mvimgnet}.}
}
\label{fig:supp_compare_MVImgeNet2}
\vspace{-0.3cm}
\end{figure*}
\begin{figure*}[tb]
\begin{center}
\includegraphics[width=0.85\linewidth]{supp_figs/compare_MVimgNet3.pdf}
\end{center}
\vspace{-6mm}
\caption{
\textbf{Qualitative comparisons with IntrinsicAny~\cite{chen2024intrinsicanything} and RGB-X~\cite{zeng2024rgb} on MVImgNet dataset~\cite{yu2023mvimgnet}.}
}
\label{fig:supp_compare_MVImgeNet3}
\vspace{-0.3cm}
\end{figure*}

\begin{figure*}[t]
\begin{center}
\vspace{0.4cm}
\includegraphics[width=0.90\linewidth]{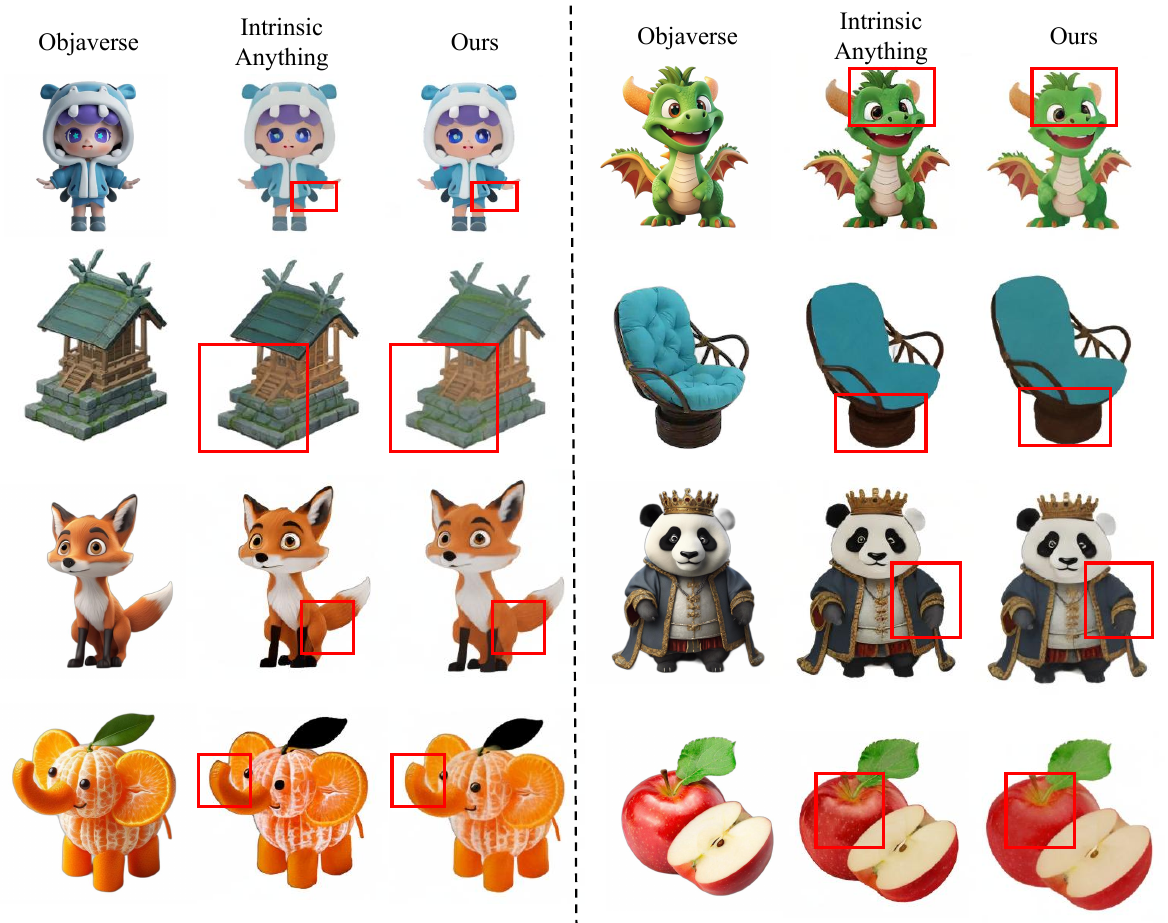}
    \end{center}
    \vspace{-0.4cm}
\caption{
\textbf{Visual comparisons with I.A.~(IntrinsicAnything) and our IR-i~(IntrinsicReal) on Objaverse~\cite{objaverse}.}
Compared to the I.A. and Ours, our contain less shadow and light information and more reasonable results.}
\label{fig:compare_syn}
\vspace{-0.1cm}
\end{figure*}

\begin{figure*}[t]
\centering
\vspace{-0.2cm}
\includegraphics[width=0.9\linewidth]{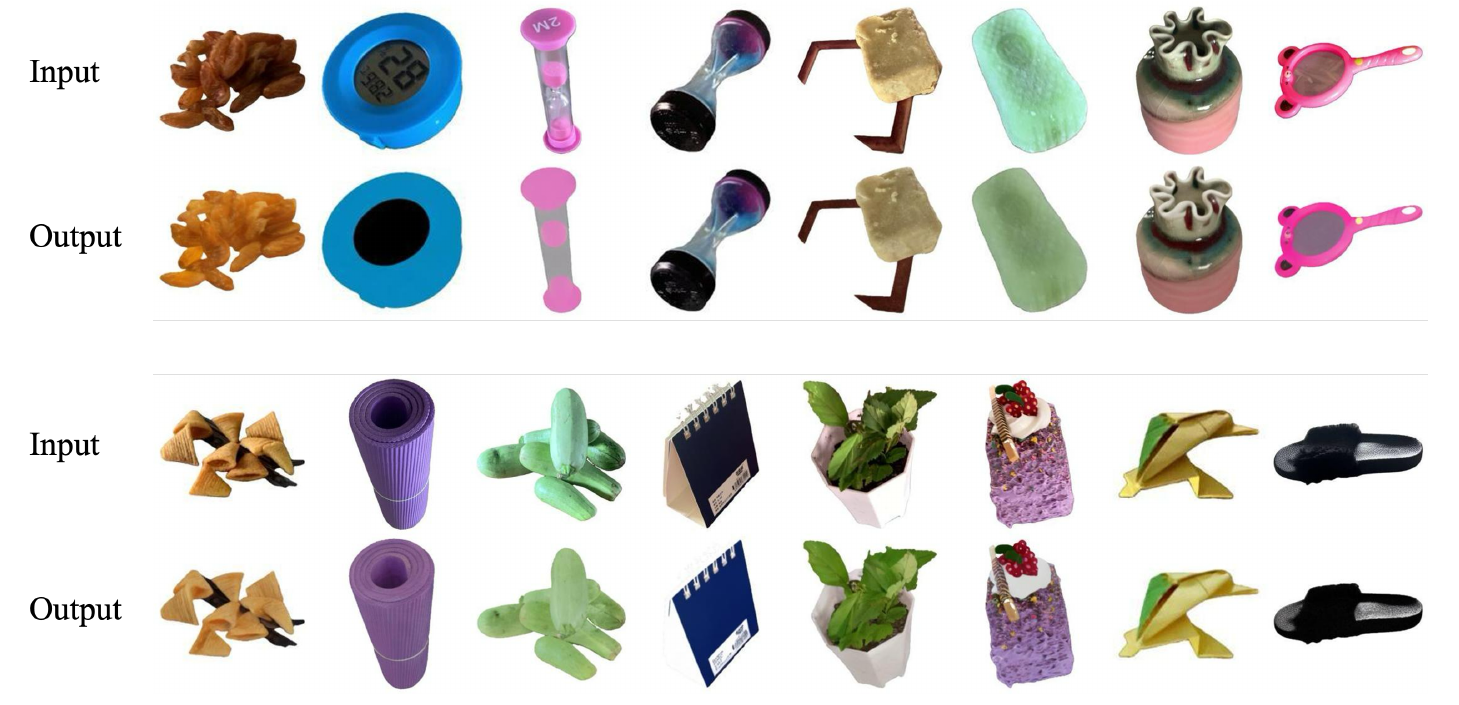}
\vspace{-0.4cm}
\caption{
\textbf{Results of applying IntrinsicAnything~\cite{chen2024intrinsicanything} on the MVImgnet dataset.}
Top row: input images. Bottom row: predicted albedos.
}
\label{fig:fig2_supp}
\vspace{-0.2cm}
\end{figure*}

\begin{figure*}[tb]
\centering
\includegraphics[width=0.97\linewidth]{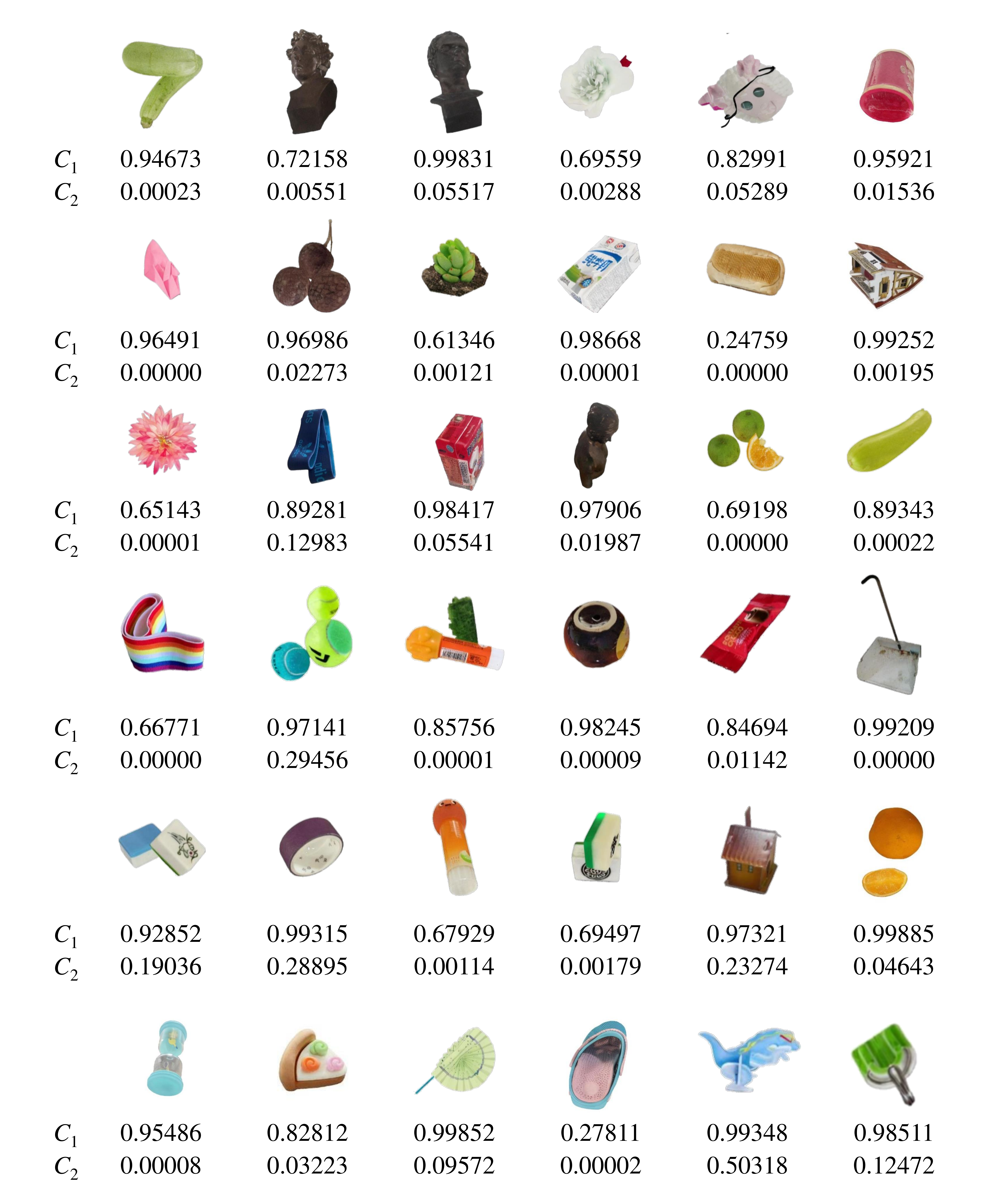}
\caption{\textbf{Scores of IR-Classifier$_1$~($C_1$) and IR-Classifier$_2$~($C_2$) on the same images.}
All the samples contain shading or lighting in this figure.  
Through iterations of the training process, the classifier's ability has been significantly improved.
For example, the cucumber in the first example has obvious highlights, which is clearly not a good albedo image. 
$C_1$ has a high probability of classifying it as an albedo image, while $C_2$ predicts it as a non albedo image.
}
\label{fig:sup_c1c2_2}
\end{figure*}

\begin{figure*}[tb]
\begin{center}
\includegraphics[width=1.0\linewidth]{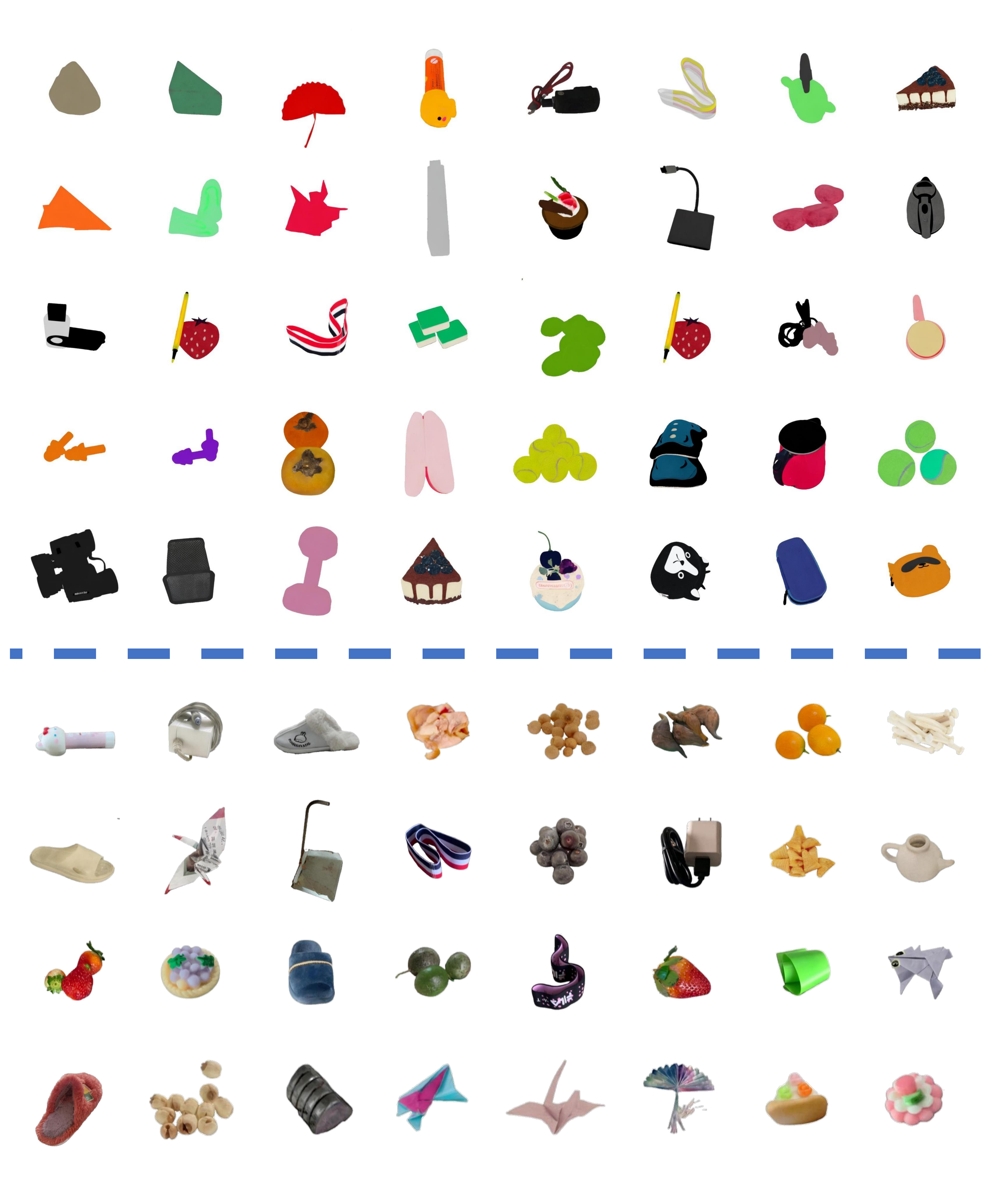}
\end{center}
\caption{\textbf{The results of IR-Classifier$_2$~($C_2$). }
The upper part of the figure visualizes results with scores above 0.99, while the lower part lists results with scores below 0.01.
}
\label{fig:sup_c1c2_1}
\end{figure*}

\begin{figure*}[t]
\begin{center}
\includegraphics[width=0.9\linewidth]{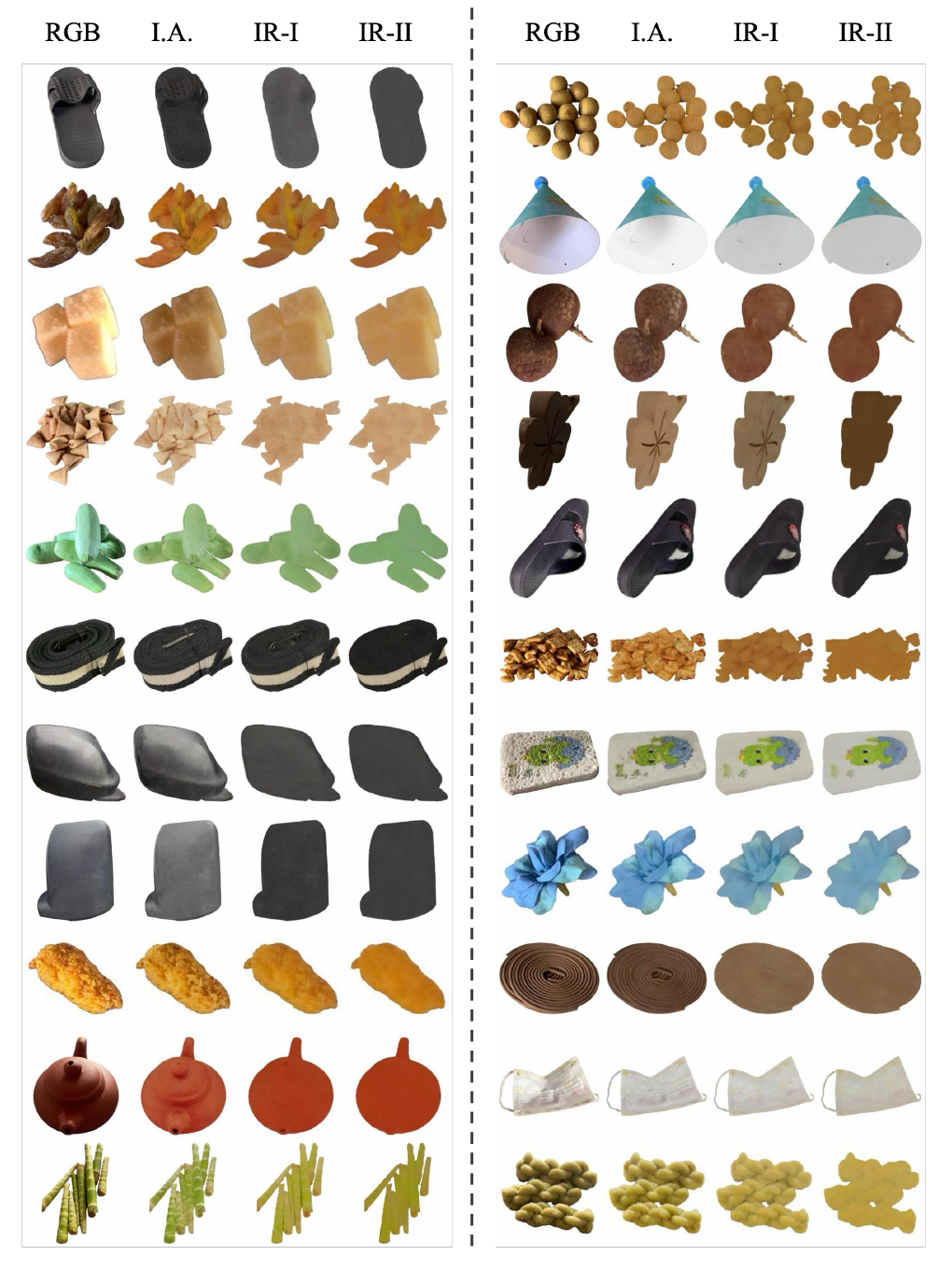}
\end{center}
\vspace{-4mm}
\caption{
\textbf{Visual comparisons with I.A.~(IntrinsicAnything) and our IR-i~(IntrinsicReal) on MVImgNet.}
From left to right, the images represent the real-world images, albedo images infer from I.A., albedo images infer from IR first iteration and second iteration, respectively.  
Compared to the I.A. and IR-i, our IR-ii contain less shadow and light information and more reasonable results.
}
\label{fig:fig7_supp}
\end{figure*}

\begin{figure*}[tb]
\begin{center}
\vspace{-0.3cm}
\includegraphics[width=0.70\linewidth]{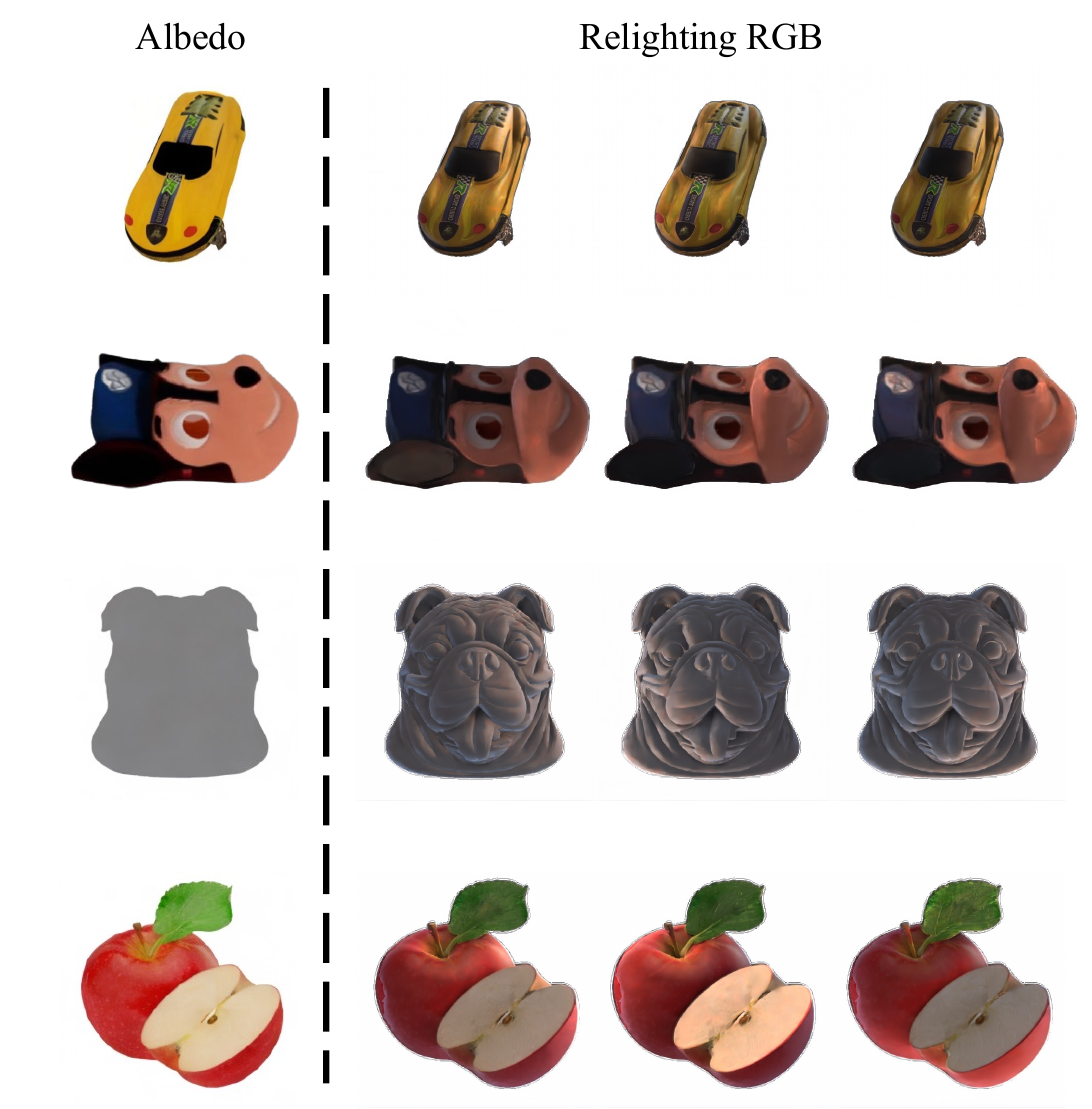}
\end{center}
\caption{\textbf{The results of relighting. }
The upper part of the figure visualizes results on the MvImgNet~\cite{yu2023mvimgnet} dataset, while the lower part lists results on the Objaverse~\cite{objaverse} dataset.
}
\label{fig:relighting}
\vspace{-0.3cm}
\end{figure*}

\begin{figure*}[t]
\begin{center}
\includegraphics[width=0.80\linewidth]{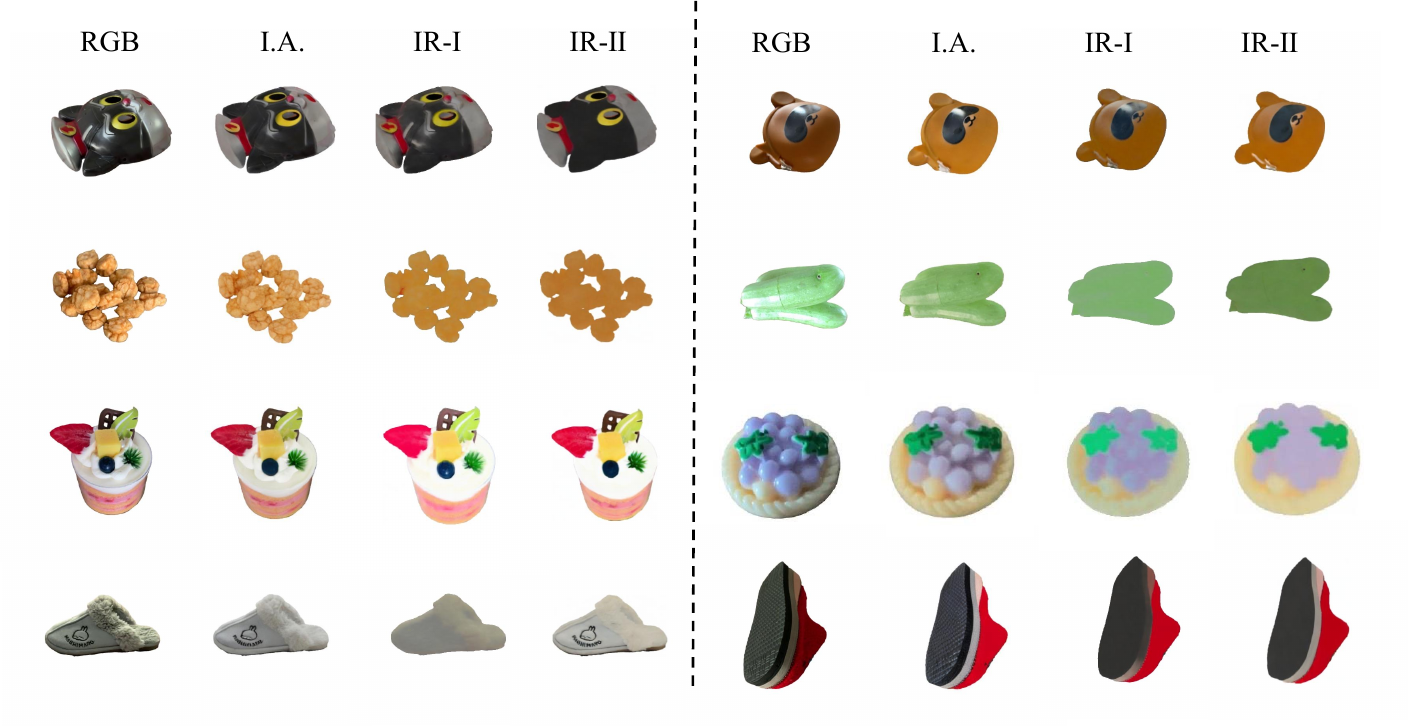}
\end{center}
\vspace{-0.5cm}
\caption{
\textbf{Visual comparisons with I.A.~(IntrinsicAnything) and our IR-i~(IntrinsicReal) on MVImgNet.}
From left to right, the images represent the real-world images, albedo images infer from I.A., albedo images infer from IR first iteration and second iteration, respectively.  
Compared to the I.A. and IR-i, our IR-ii contain less shadow and light information and more reasonable results.
}
\label{fig:fig7_supp2}
\vspace{-0.1cm}
\end{figure*}

\begin{figure*}[t]
\centering
\includegraphics[width=0.85\linewidth]{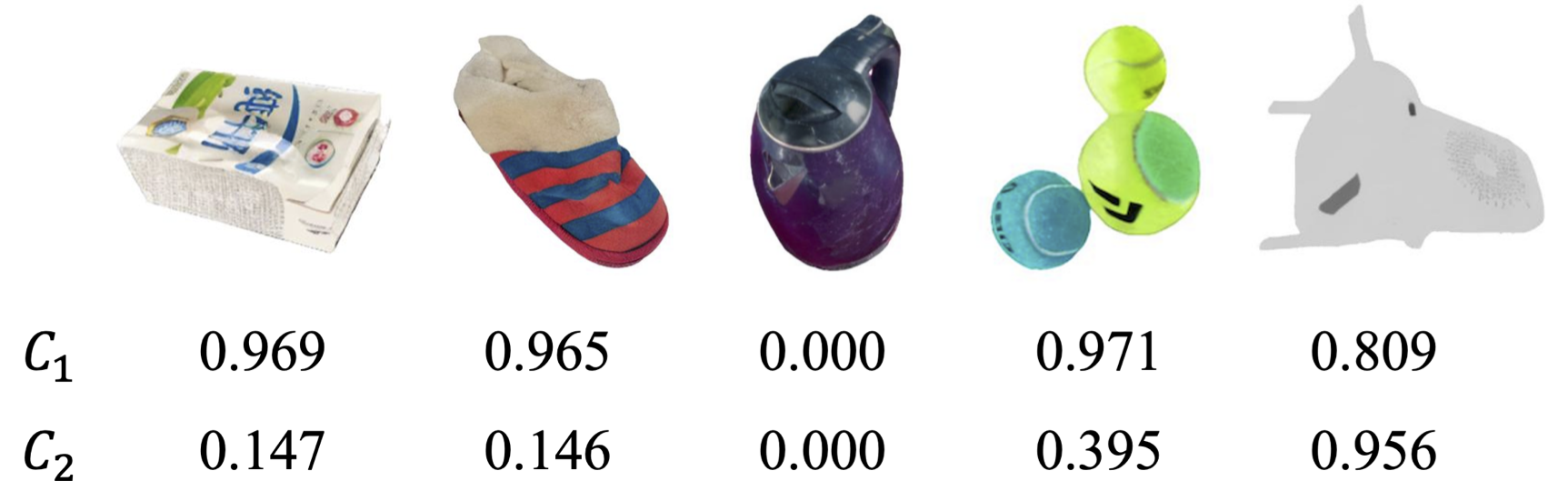}
\caption{\textbf{Scores of IR-Classifier$_1$~($C_1$) and IR-Classifier$_2$~($C_2$) on the same images.}}
\label{fig:c1c2_score}
\end{figure*}

\begin{figure*}[t]
\centering
\includegraphics[width=0.75\linewidth]{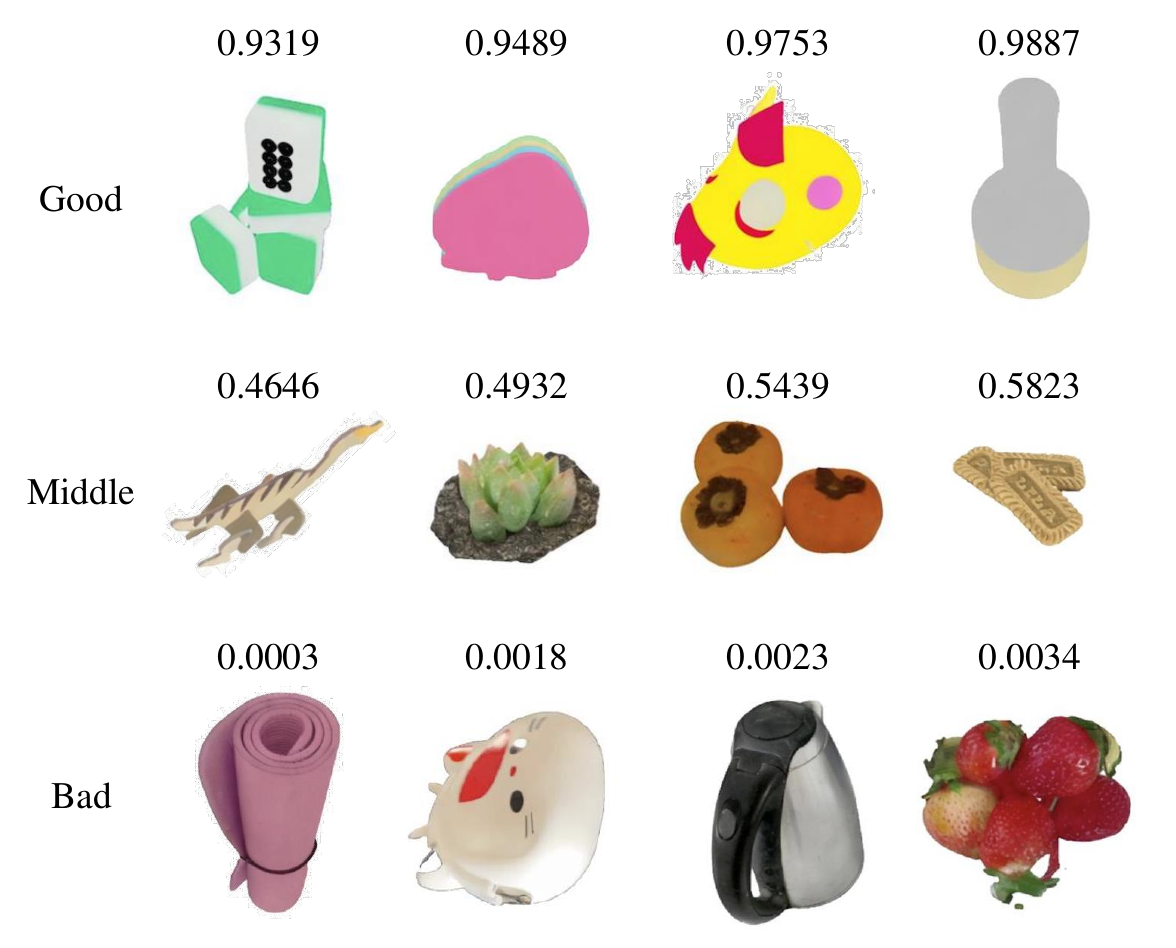}
\caption{
\textbf{Visual comparison of positive and negative results inferred from IR-Model$_0$. } 
The upper number refer to the confidence scores from IR-Classifier. 
The higher score denotes the better albedo result.}
\label{fig:compare_p_n}
\end{figure*}

\end{document}